\documentclass[11pt,a4paper]{article}
\usepackage[margin=1in]{geometry}
\usepackage[T1]{fontenc}
\usepackage{lmodern}
\usepackage{amsmath,amssymb}
\usepackage{graphicx,booktabs,array,placeins}
\usepackage[font=small,labelfont=bf]{caption}
\usepackage{microtype}
\usepackage{cite}
\usepackage[hidelinks]{hyperref}
\newsavebox{\preprinttablebox}
\newenvironment{fitwidthtable}{\begin{lrbox}{\preprinttablebox}}{\end{lrbox}\ifdim\wd\preprinttablebox>\linewidth\resizebox{\linewidth}{!}{\usebox{\preprinttablebox}}\else\usebox{\preprinttablebox}\fi}
\hypersetup{pdftitle={Angular analysis of purely baryonic Lambda-b decays},
}
\title{Angular analysis of the purely baryonic decays $\Lambda_b^0\to n p\bar p$ and $\Lambda_b^0\to\Lambda p\bar p$}

\author{
    Chao-Qiang Geng$^{1}$, Chao Han$^{1}$\\[0.5em]
    \small $^{1}$ School of Fundamental Physics and Mathematical Sciences,\\
    \small  Hangzhou Institute for Advanced Study, UCAS, Hangzhou 310024, China\\[0.5em]
    \small E-mail: \texttt{cqgeng@ucas.ac.cn}, \texttt{hbarc@ucas.ac.cn}
}

\date{}

\begin{document}
\maketitle

\begin{abstract}
An angular analysis of the purely baryonic decays $\Lambda_b^0\to np\bar p$ and $\Lambda_b^0\to\Lambda p\bar p$ is presented within a phenomenological generalized-factorization model. 
Within this model, two current contributions leading to the same final state are combined coherently. 
Pair-angle distributions, forward-backward hemisphere asymmetries, and Legendre moments are calculated in both channels, together with the polarization of the produced $\Lambda$ for an unpolarized parent. 
For the neutron channel, the full-region forward-backward asymmetries are $4.46\times10^{-4}$ in the $n\bar p$ frame and $0.856$ in the $p\bar p$ frame at $N_c^{\mathrm{eff}}=3$ and zero additional relative phase. 
From mass-bin integration, the small first asymmetry is found to arise from compensation between signed contributions from different regions. 
In the $\Lambda$ channel, the angular distributions are modified by the selection $m(p\bar p)<2.85\,\mathrm{GeV}$, and the rate-weighted longitudinal polarization along the $\Lambda$ flight direction is shifted from $-0.803$ to $-0.954$. 
These results show how angular and spin observables depend on the relative current contributions and the selected phase-space region.
The model-dependent analysis is intended as a first step toward understanding the dynamics of these decays and as a basis for further theoretical studies and comparisons with measurements.
\end{abstract}
\FloatBarrier
\section{Introduction}
\label{sec:1}

The study of weak hadron decays is important for understanding flavor-changing processes, CP violation, and the role of strong interactions in particle decays. 
Nonleptonic $B$-meson decays have been extensively explored over the past decades. 
In recent years, weak decays of bottom baryons have also attracted growing theoretical and experimental interest. 
Purely baryonic decays of bottom baryons provide a complementary setting for studying weak decay dynamics and hadronic interactions.
The three-quark structure and spin of the initial bottom baryon lead to transition dynamics that differ from those of $B$ mesons.
Their angular distributions and baryon polarizations can provide additional information about these processes beyond their integrated rates.
Of particular interest here are charmless purely baryonic three-body decays, whose final states consist exclusively of baryons and antibaryons. 
Previous studies of $\Lambda_b^0\to np\bar p$~\cite{GengHsiaoRodrigues2016} and $\Lambda_b^0\to\Lambda p\bar p$~\cite{HsiaoGengRodrigues2019} have provided predictions for branching fractions and invariant-mass spectra and, in the $\Lambda$ channel, a reference for direct CP violation. 
The rate and pair-mass spectra of the neutron mode have also been studied using an independent theoretical approach~\cite{GengHan2026}. 
Experimentally, $\Lambda_b^0\to\Lambda p\bar p$ has been observed by LHCb, and its branching fraction has been measured relative to that of $\Lambda_b^0\to\Lambda K^+K^-$ in a restricted hadron-pair invariant-mass region~\cite{LHCbLambdaHH2016,LHCbLambdaPPbar2026}. 
This measurement provides a setting for comparisons with theoretical predictions.

Against this background, the present study investigates how pair-frame projections, invariant-mass regions and phase-space selection shape the angular and spin distributions of $\Lambda_b^0\to np\bar p$ and $\Lambda_b^0\to\Lambda p\bar p$. The analysis examines pair-angle distributions, forward-backward hemisphere asymmetries and Legendre moments in both channels, together with the $\Lambda$ polarization for an unpolarized parent. 
Specifically, the analysis compares the two pair-rest-frame projections and examines whether a nearly vanishing integrated asymmetry can conceal compensating contributions across invariant mass. 
It also studies how phase-space cuts alter the angular and spin distributions.
For the $\Lambda$ channel, the selection $m(p\bar p)<2.85\,\mathrm{GeV}$ is considered in connection with the experimental measurement.

To address these questions, the calculation adapts the phenomenological generalized-factorization prescription originally developed for mesonic weak decays to the two baryonic channels~\cite{AliKramerLu1998}. 
In this framework, two current contributions to each final state are combined coherently, with each expressed as a product of a heavy-baryon transition current and a vacuum baryon-pair production current.
To construct the angular and spin observables transparently, these amplitudes are organized in a common external-spin representation, where current-helicity components resolve the vector and axial contractions and supply the bilinear responses required for both the unpolarized angular distributions and the daughter polarization. 

For the neutron channel, the reference inputs give a nearly vanishing neutron-antiproton hemisphere asymmetry and a strongly asymmetric proton-antiproton projection. 
Integration over invariant-mass bins shows that the small first asymmetry results from compensation between oppositely signed contributions from different mass regions. 
In the $\Lambda$ channel, imposing the $p\bar p$ invariant-mass selection significantly alters both the angular asymmetries and the regional daughter-spin composition.
Integrated rates and invariant-mass spectra are also calculated to show the overall decay strengths and the mass regions relevant to the angular and spin distributions. 
The same coherent amplitudes are used to calculate complementary direct CP asymmetries and a selected CP-averaged branching fraction for comparison with experiment~\cite{LHCbLambdaHH2016,LHCbLambdaPPbar2026}. 
These findings illustrate, within the adopted model, how angular and spin observables depend on the current contributions and the selected phase-space region.

The remainder of the paper is organized as follows. 
The phenomenological decay amplitudes and the common-spin density are developed in Sections~\ref{sec:2} and~\ref{sec:3}. 
The three-body kinematics, angular observables, and polarization are then defined in Section~\ref{sec:4}. 
Section~\ref{sec:5} brings together the numerical inputs and uncertainty scope, the angular and spin predictions in their rate and mass-spectrum context, and the complementary CP and experimental comparisons. 
Conclusions are given in Section~\ref{sec:6}.
Appendix A collects the operator conventions and input specifications, while Appendix B gives the tensor representation and kernel construction.

\FloatBarrier

\section{Phenomenological framework and baryonic decay amplitudes}
\label{sec:2}

Within the adopted model, the angular distributions and daughter polarization depend on two transition-pair current contributions and their interference. 
To calculate these observables, the relative weights of the contributions and their hadronic matrix elements must first be specified. 
Section~\ref{sec:2-1} introduces the decay conventions and effective weak coefficients. 
The coherent combination of the two contributions
is described in Section~\ref{sec:2-2}. 
The hadronic currents and densities are introduced in Section~\ref{sec:2-3}.

\FloatBarrier
\subsection{Decay conventions and effective Hamiltonian}
\label{sec:2-1}

The two decays are described as $\Lambda_b^0(P)\to \mathbf{B}_q(p_1)p(p_2)\bar p(p_3)$, where $\mathbf{B}_d=n$, $\mathbf{B}_s=\Lambda$, and $q=d,s$ labels the weak transition. 
In the first, the $\Lambda_b^0$ undergoes a transition to a proton and the $\mathbf{B}_q\bar p$ pair is produced from the vacuum. 
In the second, the transition produces $\mathbf{B}_q$ and the $p\bar p$ pair is produced from the vacuum. 
These contributions are illustrated for the neutron channel in Fig.~\ref{fig:current-assignments}.
The first contribution is labeled $13$, according to the particles in the produced pair, and the second is labeled $23$. 
Their momentum transfers are $q_{13}=P-p_2=p_1+p_3$ and $q_{23}=P-p_1=p_2+p_3$, respectively. 
The corresponding pair-current invariants are $t_a=q_a^2=s_a$ for $a=13,23$. 
Both contributions are expressed with the same external-spin ordering: parent, $\mathbf{B}_q$, proton, and antiproton.

In the generalized-factorization prescription, baryonic current products are combined with process-dependent effective coefficients~\cite{AliKramerLu1998,GengHsiaoRodrigues2016,GengLiuTsai2021}.
An effective color parameter $N_c^{\mathrm{eff}}$ is introduced to model nonfactorizable effects in the color weights.
The effective weak Hamiltonian supplies the operator coefficients entering these current products. 
It includes current-current operators and four-quark QCD and electroweak penguins~\cite{GengLiuTsai2021}. 
With the CKM products $\lambda_r^{(q)}=V_{rb}V_{rq}^*$, the Hamiltonian is
\begin{equation}
\label{eq:unified-hamiltonian}
\mathcal H_{\mathrm{eff}}^{(q)}
=\frac{G_F}{\sqrt2}
\left[\sum_{r=u,c}\lambda_r^{(q)}(C_1O_1^{r,q}+C_2O_2^{r,q})
-\lambda_t^{(q)}\sum_{i=3}^{10}C_iO_i^{(q)}\right]
+\mathrm{h.c.}
\end{equation}
The operators $O_{3\text{-}6}$ and $O_{7\text{-}10}$ describe QCD and four-quark electroweak penguins, respectively. 
The electromagnetic and chromomagnetic dipole operators $O_{7\gamma}$ and $O_{8g}$ are omitted because their baryonic matrix elements are not evaluated here. 
Direct charmless current-current production selects $r=u$, while short-distance penguin contractions enter the process-dependent effective coefficients $c_i^{\mathrm{eff}}$, which differ from the renormalized Hamiltonian coefficients $C_i$.

To obtain the weights of the factorized current products, the paired four-quark operators are projected onto color-singlet products. 
This projection generates direct and inverse-color terms. 
In the present prescription, $1/N_c$ is replaced by $\xi=1/N_c^{\mathrm{eff}}$, giving
\begin{equation}
\label{eq:unified-color-combinations}
a_{2j-1}=c_{2j-1}^{\mathrm{eff}}+\xi c_{2j}^{\mathrm{eff}},\qquad
a_{2j}=c_{2j}^{\mathrm{eff}}+\xi c_{2j-1}^{\mathrm{eff}},\qquad
\xi=\frac1{N_c^{\mathrm{eff}}},\quad j=1,\ldots,5.
\end{equation}
The neutron and $\Lambda$ modes use the $b\to d$ and $b\to s$ effective coefficients, respectively~\cite{GengLiuTsai2021}. 
A common real $N_c^{\mathrm{eff}}$ is assumed for all $a_i$ and both channels.
Its variation probes sensitivity to the adopted color weights at fixed hadronic inputs. 
For each current assignment, the operator contributions can be grouped into tree, QCD-penguin, and electroweak-penguin terms:
\begin{equation}
\label{eq:full-assignment-operator-content}
\mathcal M_a^{(q)}
=\mathcal M_{a,T}^{(q)}
+\mathcal M_{a,P_{\mathrm{QCD}}}^{(q)}
+\mathcal M_{a,P_{\mathrm{EW}}}^{(q)},
\qquad a=13,23.
\end{equation}
The penguin terms include the scalar-pseudoscalar density products defined in Section~\ref{sec:2-3}.

\FloatBarrier
\subsection{Coherent current assignments and interference}
\label{sec:2-2}

\begin{figure}[tbp]
\centering
\includegraphics[width=\linewidth,height=0.72\textheight,keepaspectratio]{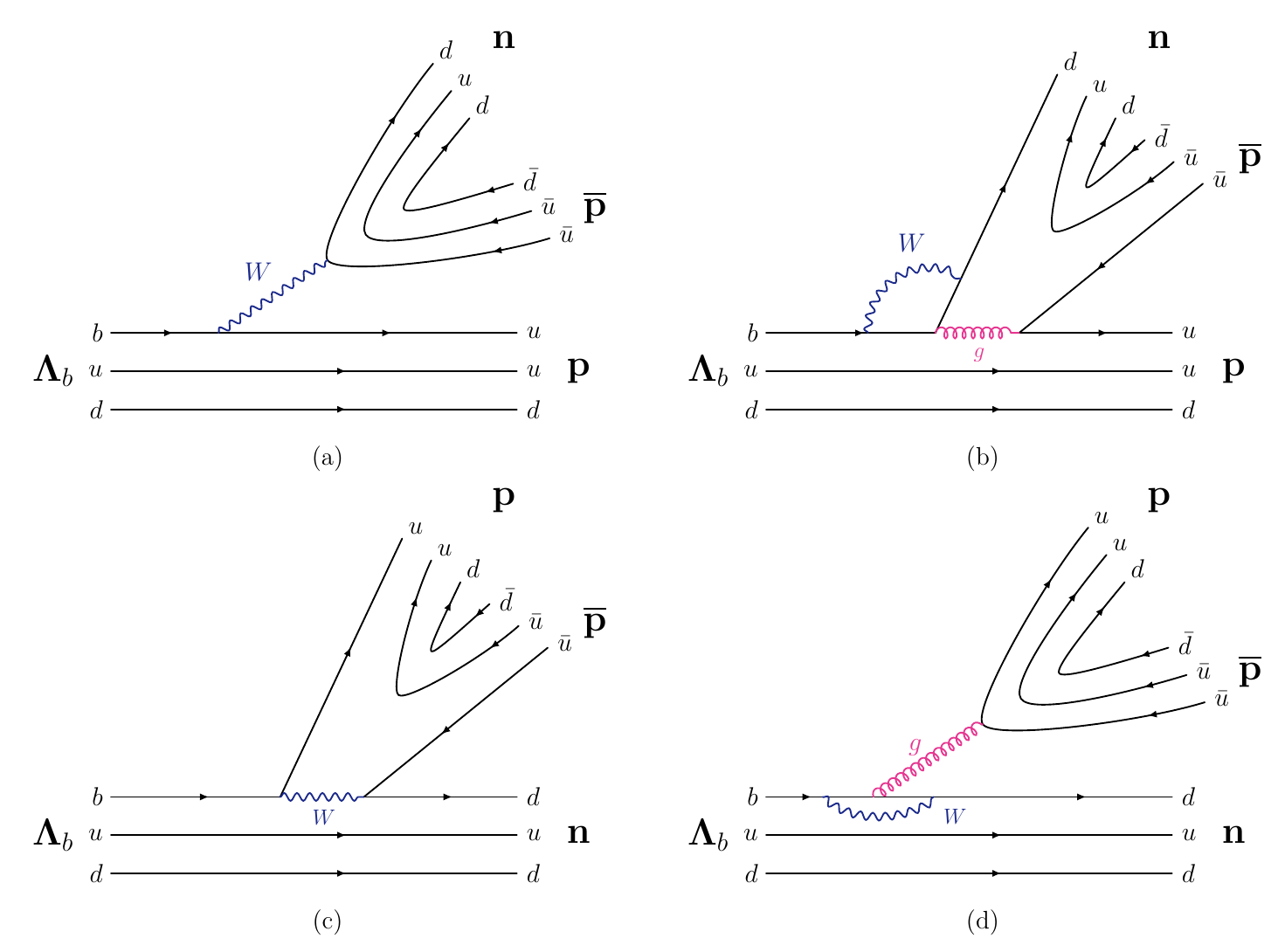}
\caption{
Quark-level topologies for $\Lambda_b^0\to np\bar p$ are shown for tree diagrams in panels (a) and (c) and for QCD-penguin diagrams in panels (b) and (d). 
The diagrams are adapted from Ref.~\cite{GengHsiaoRodrigues2016}. Panels (a) and (b) contain a proton transition with $n\bar p$ production and correspond to assignment $13$. 
Panels (c) and (d) contain a neutron transition with $p\bar p$ production and correspond to assignment $23$. 
The figure illustrates the neutron channel, while the current products for both channels are defined in the text.
}
\label{fig:current-assignments}
\end{figure}

Specifically, Fig.~\ref{fig:current-assignments} illustrates the tree and QCD-penguin topology classes of the neutron channel~\cite{GengHsiaoRodrigues2016}. 
Panels (a) and (b) contain a proton transition with $n\bar p$ production, whereas panels (c) and (d) contain a neutron transition with $p\bar p$ production.

The two current products enter with different tree-level color weights. 
To specify these weights, the tree-operator matrix elements are written in terms of the left-handed products $\mathcal F_{13}^{(q)}$ and $\mathcal F_{23}^{(q)}$ through a phenomenological matching prescription. 
In the reference phase convention, this prescription gives
\begin{equation}
\label{eq:unified-topology-matching}
\begin{pmatrix}\langle O_1^{u,q}\rangle\\\langle O_2^{u,q}\rangle\end{pmatrix}
\simeq
\begin{pmatrix}1&\xi\\\xi&1\end{pmatrix}
\begin{pmatrix}\mathcal F_{13}^{(q)}\\\mathcal F_{23}^{(q)}\end{pmatrix}.
\end{equation}
Before the additional relative phase and apart from the common weak normalization, the prescription gives the tree weights $\lambda_u^{(q)}a_1$ and $\lambda_u^{(q)}a_2$ for the two products. 
The complete assignments also contain the penguin contributions introduced in Section~2.1 and specified in Section~2.3.
Because the two contributions lead to the same external final state, their complete amplitudes are combined coherently: 
\begin{equation}
\label{eq:unified-coherent-ansatz}
\mathcal M^{(q)}=\mathcal M_{13}^{(q)}+e^{i\delta_q}\mathcal M_{23}^{(q)}.
\end{equation}
The two current products are combined coherently within the adopted phenomenological prescription. 
Their relative weights are specified by the matching scheme introduced above.
Nonfactorizable effects beyond those represented by the effective color parameter are not calculated explicitly.
The baseline amplitudes contain CKM weak phases and absorptive penguin contributions through the process-dependent coefficients $c_i^{\mathrm{eff}}$~\cite{GengLiuTsai2021}. 
The form factors are taken to be real. 
An additional CP-even phase $\delta_q$ is introduced between the two current contributions and is assumed to be constant across phase space for each channel. 
Its variation probes a relative hadronic phase without modeling momentum-dependent final-state interactions. 
The weak and absorptive phases remain present when $\delta_q=0$.

\FloatBarrier
\subsection{Hadronic currents and form factors}
\label{sec:2-3}

The factorized amplitudes are constructed from heavy-baryon transition and vacuum pair-production matrix elements. 
The corresponding vector and axial currents are defined first. 
Scalar and pseudoscalar densities are then introduced for the left-right penguin contributions, after which the two amplitudes and their flavor coefficients are assembled.
For a transition to $N=p,n,\Lambda$, define $Q=P-p_N$, $t=Q^2$, $M=m_{\Lambda_b}$, and the quark flavors $q_p=u$, $q_n=d$, $q_\Lambda=s$. 
In either current assignment, $Q$ is the momentum carried by the associated baryon-antibaryon pair. 
The transition current is $J_N^\mu=\langle N|\bar q_N\gamma^\mu(1-\gamma_5)b|\Lambda_b\rangle=V_N^\mu-A_N^\mu$, with
\begin{equation}
\label{eq:unified-transition-currents}
\begin{aligned}
V_N^\mu&=\bar u_N\left[f_1^N\gamma^\mu
+\frac{f_2^N}{M}i\sigma^{\mu\nu}Q_\nu
+\frac{f_3^N}{M}Q^\mu\right]u_{\Lambda_b},\\
A_N^\mu&=\bar u_N\left[g_1^N\gamma^\mu
+\frac{g_2^N}{M}i\sigma^{\mu\nu}Q_\nu
+\frac{g_3^N}{M}Q^\mu\right]\gamma_5u_{\Lambda_b}.
\end{aligned}
\end{equation}
The conventions are $g^{\mu\nu}=\operatorname{diag}(1,-1,-1,-1)$ and $\sigma^{\mu\nu}=i[\gamma^\mu,\gamma^\nu]/2$. 
The covariant transition form factors $f_i^N(t)$ and $g_i^N(t)$ depend on the current invariant $t$. 
The positive tensor sign fixes their conversion from the lattice helicity functions in Appendix A.

For vacuum production of $\mathbf{B}_i \mathbf{\bar B}_j$, define the pair momentum $Q=p_i+p_j$ and mass sum $\Sigma_{ij}=m_i+m_j$. 
The flavor label $x=q_1q_2$ identifies the current $\bar q_1\Gamma q_2$ and its external pair. 
The adopted pair currents contain four form factors:
\begin{equation}
\label{eq:unified-pair-currents}
\begin{aligned}
V_\mu^x&=\bar u_i\left[F_1^x\gamma_\mu
+\frac{F_2^x}{\Sigma_{ij}}i\sigma_{\mu\nu}Q^\nu\right]v_j,\quad
A_\mu^x&=\bar u_i\left[g_A^x\gamma_\mu
+\frac{h_A^x}{\Sigma_{ij}}Q_\mu\right]\gamma_5v_j.
\end{aligned}
\end{equation}
The charged pair currents $V_\mu^{qu}$ and $A_\mu^{qu}$ produce $\mathbf{B}_q\bar p$, while the neutral currents $V_\mu^{ff}$ and $A_\mu^{ff}$ produce $p\bar p$. 
For the neutral currents, the superscript $ff$ identifies the quark flavor of each current. 
The vector form factors $F_1^{ff}$ and $F_2^{ff}$ therefore describe individual flavor-current contributions, whereas the proton electromagnetic form factors involve charge-weighted sums of these contributions.
The axial form factors $g_A^{ff}$ and $h_A^{ff}$ describe the corresponding flavor-specific axial currents.
The pair currents follow the four-form-factor parameterization of Ref.~\cite{HuangHsiaoWangSun2022}. 
The left-right penguin operators also generate scalar and pseudoscalar density products through a Fierz rearrangement. 
The corresponding density form factors are defined by
\begin{equation}
\label{eq:unified-density-definitions}
\begin{aligned}
\langle N|\bar q_N b|\Lambda_b\rangle&=f_S^N\bar u_Nu_{\Lambda_b},&
\langle N|\bar q_N\gamma_5b|\Lambda_b\rangle&=g_P^N\bar u_N\gamma_5u_{\Lambda_b},\\
\langle \mathbf{B}_i\mathbf{\bar B}_j|\bar q_1q_2|0\rangle&=F_S^x\bar u_iv_j,&
\langle \mathbf{B}_i \mathbf{\bar B}_j|\bar q_1\gamma_5q_2|0\rangle&=G_P^x\bar u_i\gamma_5v_j.
\end{aligned}
\end{equation}
For convenience, the scalar and pseudoscalar matrix elements are combined as $S_N^-=\langle N|\bar q_N(1-\gamma_5)b|\Lambda_b\rangle$ for the heavy-baryon transition and $D_x^+=\langle \mathbf{B}_i  \mathbf{\bar B}_j|\bar q_1(1+\gamma_5)q_2|0\rangle$ for baryon-antibaryon pair production from the vacuum.
The charged-pair contribution is then given by
\begin{equation}
\label{eq:unified-charged-amplitude}
\begin{aligned}
\mathcal M_{13}^{(q)}
&=\frac{G_F}{\sqrt2}\left[
\kappa_q^{VA}J_p^\mu(V_\mu^{qu}-A_\mu^{qu})
+\kappa_q^{SP}S_p^-D_{qu}^+\right],\\
\kappa_q^{VA}&=\lambda_u^{(q)}a_1-\lambda_t^{(q)}(a_4+a_{10}), \quad
\kappa_q^{SP}=2\lambda_t^{(q)}(a_6+a_8).
\end{aligned}
\end{equation}
The neutral-current contribution is summed over the pair flavors $f=u,d,s$, with their tree and penguin weights kept explicit:
\begin{equation}
\label{eq:unified-neutral-amplitude}
\mathcal M_{23}^{(q)}
=\frac{G_F}{\sqrt2}\left[
J_{\mathbf{B}_q}^{\mu}\sum_{f=u,d,s}
\left(\beta_{f,q}^+V_\mu^{ff}-\beta_{f,q}^-A_\mu^{ff}\right)
+S_{\mathbf{B}_q}^-\sum_{f=u,d,s}\eta_{f,q}D_{ff}^+\right],
\end{equation}
where
\begin{equation}
\label{eq:unified-neutral-flavor-map}
\begin{aligned}
\beta_{f,q}^{\pm}
&=\lambda_u^{(q)}a_2\delta_{fu}
-\lambda_t^{(q)}\left[
a_3\pm a_5+\frac32e_f(a_9\pm a_7)
+\delta_{fq}\left(a_4+\frac32e_fa_{10}\right)\right],\\
\eta_{f,q}
&=2\lambda_t^{(q)}\delta_{fq}\left(a_6+\frac32e_fa_8\right),
\qquad e_u=\frac23,\quad e_d=e_s=-\frac13.
\end{aligned}
\end{equation}
The flavor coefficients combine the up-current tree contribution, direct flavor-diagonal penguins, and flavor-matched terms in the conventions of Refs.~\cite{HsiaoGengYuZhao2019,HsiaoTsaiRodrigues2020}. 
For the Dirac structures considered here, the reference amplitudes use the valence/OZI approximation $\langle p\bar p|\bar s\Gamma s|0\rangle=0$~\cite{Iizuka1966}. 
The accuracy of this approximation in the timelike region, however, is not quantified.
The coefficient $\eta_{f,q}$ selects the neutral down density in the neutron channel and the neutral strange density in the $\Lambda$ channel. 
The adopted strange-current omission removes the latter, while the charged $\Lambda\bar p$ density and the neutron neutral down-density contributions remain.

The scalar and pseudoscalar contributions require form factors for both the heavy-baryon transition and baryon-antibaryon pair production. 
For the transition matrix elements, the scalar and pseudoscalar form factors are obtained from the vector and axial-vector currents through the quark equations of motion. 
Using consistent quark masses and renormalization conventions, the reference transition form factors are
\begin{equation}
\label{eq:unified-transition-ward}
f_S^N(t)=\frac{(M-m_N)f_1^N(t)+(t/M)f_3^N(t)}{m_b-m_{q_N}},
\qquad
g_P^N(t)=\frac{(M+m_N)g_1^N(t)-(t/M)g_3^N(t)}{m_b+m_{q_N}}.
\end{equation}
For the timelike pair densities $F_S^x(t)$ and $G_P^x(t)$, the reference calculation uses the functions from the global fit of Huang et al.~\cite{HuangHsiaoWangSun2022} and the flavor relations in Appendix A. 

\FloatBarrier
\section{Spin structure and bilinear responses}
\label{sec:3}

The amplitudes of Section~\ref{sec:2} must be expressed in common external spin states before the two contributions can be combined consistently and their interference evaluated. 
Section~\ref{sec:3-1} first resolves the vector-axial contractions within each assignment into current-helicity components, then relates the external states of the two assignments to a common spin basis. 
Averaging over the parent spin and summing the final spins gives the scalar intensity, separated in Section~\ref{sec:3-2} into isolated contributions and a phase-dependent coherent overlap. 
The daughter-spin information is preserved in Section~\ref{sec:3-3} by leaving its spin indices open for a specified parent density. 
The resulting density is expressed through physical Pauli responses and equivalent coefficient kernels in Section~\ref{sec:3-4}, preparing the polarization observables of Section~\ref{sec:4}.

\FloatBarrier
\subsection{Common external spin states and current polarizations}
\label{sec:3-1}

The current-helicity basis resolves the Lorentz contraction separately within each assignment. 
For the final baryon $Y=\mathbf{B}_q$ and additional relative strong phase $\delta_Y=\delta_q$, assignment $a=13$ or $23$ identifies the produced $Y\bar p$ or $p\bar p$ pair. 
Its current momentum is $q_a$, with $q_a^2=s_a>0$. 
To resolve the contraction of the transition and pair currents, a polarization basis is introduced for the timelike momentum $q_a$.
The labels $\lambda=t,0,+1,-1$ denote the timelike, longitudinal, and two transverse current components. 
The basis consists of the timelike vector $\epsilon_t^\mu=q_a^\mu/\sqrt{s_a}$ and three orthonormal spacelike vectors $\epsilon_\lambda^\mu$, with $\lambda=0,\pm1$, satisfying
\begin{equation}
\label{eq:polarization-completeness}
g^{\mu\nu}
=\epsilon_t^\mu\epsilon_t^{*\nu}
-\sum_{\lambda=0,\pm1}\epsilon_\lambda^\mu\epsilon_\lambda^{*\nu}.
\end{equation}
For a transition current $J^\mu$ and pair current $B^\mu$, the polarization contractions are $\mathcal H_\lambda=J_\mu\epsilon_\lambda^{*\mu}$ and $\mathcal L_\lambda=\epsilon_\lambda^\nu B_\nu$. At fixed external spins,

\begin{equation}
\label{eq:current-helicity-product}
J_\mu B^\mu
=\mathcal H_t\mathcal L_t
-\mathcal H_0\mathcal L_0
-\mathcal H_{+1}\mathcal L_{+1}
-\mathcal H_{-1}\mathcal L_{-1}.
\end{equation}
This exact resolution~\cite{KadeerKornerMoosbrugger2009} identifies the timelike product $\mathcal H_t\mathcal L_t=(q_a\cdot J)(q_a\cdot B)/\allowbreak s_a$ as the current-divergence part of the vector-axial contraction. The timelike vector $\epsilon_t^\mu$ follows $q_a$ rather than a fixed parent-frame time axis. 
The scalar-pseudoscalar penguin products enter separately in the amplitude with coefficients $\kappa_q^{SP}$ and $\eta_{f,q}$.

Resolving each current contraction does not by itself ensure that the external-spin labels in the two assignments represent the same physical states. 
Helicity states defined through different pair rest frames generally differ by rotations and particle-antiparticle state phases, which must be included when the amplitudes are expressed in a common basis~\cite{Marangotto2020}. 
For assignment $a$, the reduced current-helicity couplings are embedded in the complete external-spin space as the current-product matrix $Y_a$. 
The transformation $U_a$ then maps them to the common-spin matrix $X_a$ through
\begin{equation}
\label{eq:helicity-transport}
X_a=U_aY_a,\qquad U_a^\dagger U_a=\mathbf1,\qquad
X_{13}^\dagger X_{23}
=Y_{13}^\dagger U_{13}^\dagger U_{23}Y_{23}.
\end{equation}
The convention- and kinematics-dependent phases carried by $U_a$ are distinct from the additional relative strong phase $\delta_Y$ of Section~\ref{sec:2-2}.
In the calculation, both contributions are evaluated with the same external Dirac spinors, which expresses them in a common spin basis. 
Appendix B gives the explicit current-product columns and connected-trace checks. 

\FloatBarrier
\subsection{Scalar contractions and Gram matrix $K$}
\label{sec:3-2}

The unpolarized momentum distributions are obtained from the intensity averaged over the parent spin and summed over the final spins. 
To account for the interference between the two assignments in this sum, the coherent amplitude is written as a combination of dynamical coefficients and kinematic spinor products.

For each assignment, the reference coefficient vector $H_a^{(Y,0)}$ collects the weak and form-factor weights. 
It includes the common normalization $G_F/\sqrt2$, the CKM factors, and the absorptive contributions in the adopted effective coefficients.
Only the additional relative phase is left explicit. 
The common-spin matrix $X^{(Y)}$ contains the remaining kinematic spinor products. 
The channel label $Y=\mathbf{B}_q$ denotes the daughter baryon and is distinct from the current-product matrix $Y_a$ in the common-spin transformation. 
Their combination gives
\begin{equation}
\label{eq:coherent-spin-amplitude}
\mathcal M_h^{(Y)}
=\sum_I X^{(Y)}_{h,13;I}H^{(Y,0)}_{13;I}
+e^{i\delta_Y}\sum_J X^{(Y)}_{h,23;J}H^{(Y,0)}_{23;J}
=\sum_I X^{(Y)}_{hI}H_I^{(Y)}.
\end{equation}
Each row $h$ specifies the parent and three final-particle spins in the common basis. 
The columns $I,J$ label the transition-pair Lorentz products listed in Appendix B, including vector-axial and scalar-pseudoscalar structures. 
The combined vector $H^{(Y)}$ contains both coefficient blocks, with the second multiplied by $e^{i\delta_Y}$.
The coefficients include the weak normalization, so no additional factor is required.
Because the current-product columns need not be orthogonal or independent, the individual quantities $|H_I|^2$ cannot generally be interpreted as probabilities for physical helicity states.
Summing their products over spins gives the Gram matrix $K$ and hence the intensity with one parent-spin average:
\begin{equation}
\label{eq:unified-gram-master}
\overline{|\mathcal M|^2}
=\frac12\sum_h|\mathcal M_h|^2
=H^\dagger K H,
\qquad K=\frac12X^\dagger X.
\end{equation}
The matrix $K$ connects the current-product coefficients to the spin-averaged intensity. 
Its block structure separates the two assignment contributions and their interference.
Grouping the columns of $X$ by assignment partitions $K$ into the diagonal blocks $K_{13,13}$ and $K_{23,23}$, which give the isolated intensities while retaining interference among their own current products. 
The cross block $K_{13,23}$ contains the allowed vector, axial, scalar and pseudoscalar overlaps between assignments and gives the coherent interference intensity $W_{\mathrm{int}}$,
\begin{equation}
\label{eq:unified-kernel-blocks}
K=
\begin{pmatrix}
K_{13,13}&K_{13,23}\\
K_{13,23}^{\dagger}&K_{23,23}
\end{pmatrix},
\qquad
W_{\mathrm{int}}
=2\operatorname{Re}(H_{13}^\dagger K_{13,23}H_{23}).
\end{equation}
To examine the effect of the relative phase on the integrated rate, the two assignment contributions and their interference are integrated over the same phase-space region.
The isolated intensities give $\mathcal B_{13,Y}$ and $\mathcal B_{23,Y}$.
Since $\delta_Y$ is constant over phase space, its factor $e^{i\delta_Y}$ can be taken outside the complex overlap integral of the amplitudes before the additional relative phase is applied.
Including the phase-space and lifetime normalization defines $\mathcal I_Y$ and gives
\begin{equation}
\label{eq:integrated-overlap}
\begin{aligned}
\mathcal I_Y&=
{\tau_{\Lambda_b}}
\int\frac{d\Phi_3}{2M}\,
\frac12\sum_h\mathcal M_{13,h}^{(Y)*}\mathcal M_{23,h}^{(Y)},\\
\mathcal B_Y(\delta_Y)
&=\mathcal B_{13,Y}+\mathcal B_{23,Y}
+2\operatorname{Re}(e^{i\delta_Y}\mathcal I_Y),\quad
|\mathcal I_Y|^2&\le\mathcal B_{13,Y}\mathcal B_{23,Y}.
\end{aligned}
\end{equation}
At fixed weak coefficients and form factors, the additional relative phase affects the integrated branching fraction through the interference term, while the two isolated contributions and their incoherent sum remain unchanged. 
The coherent rate can therefore be enhanced or suppressed relative to the incoherent sum, and the normalized angular distributions can also change.
The overlap $\mathcal I_Y$ is calculated from the amplitudes before the additional relative phase is applied, so its phase $\arg\mathcal I_Y$ is not an independent fit parameter.
Although the overlap bound limits the magnitude of the interference, its contribution can exceed the isolated rate of the smaller amplitude.

\FloatBarrier
\subsection{Parent and daughter spin densities}
\label{sec:3-3}

The spin-summed intensity determines the unpolarized momentum distributions but does not retain the daughter-spin information needed to calculate polarization.
This information is preserved by leaving the two daughter-spin indices open in the product of the coherent amplitude and its complex conjugate, while summing the unobserved proton and antiproton spins. The construction is given below for $\Lambda$ and applies formally to the neutron as well.

The daughter-spin density depends on the initial spin state of the parent. 
In the common parent-spin basis, this state is described by the density matrix $\rho_b$, which can be written in terms of the parent polarization vector $\mathbf P_b$~\cite{Marangotto2020} as
\begin{equation}
\label{eq:spin-parent-density}
\rho_b=\frac{\mathbf1+\mathbf P_b\cdot\boldsymbol\sigma^{(b)}}2,
\qquad \operatorname{Tr}\rho_b=1,
\qquad |\mathbf P_b|\leq1.
\end{equation}
The Pauli matrices $\boldsymbol\sigma^{(b)}$ act on the two parent-spin states. 
The diagonal entries of the trace-one density matrix $\rho_b$ give their relative probabilities, while its off-diagonal entries describe their coherence.
For an unpolarized parent, setting $\rho_b=\mathbf1/\allowbreak 2$ includes the single parent-spin average of Eq.~(\ref{eq:unified-gram-master}).
 
The external-spin row $h$ introduced in Eq.~\eqref{eq:coherent-spin-amplitude} can be written as $h=(s_b,s_\Lambda,\eta)$, where $\eta=(s_p,s_{\bar p})$ groups the proton and antiproton spin labels.
Suppressing the $\Lambda$ channel superscript, the coherent amplitude remains $\mathcal M_{s_b s_\Lambda\eta}=\sum_I X_{s_b s_\Lambda\eta,I}H_I$. Contracting the parent indices with $\rho_b$ while leaving the daughter indices open gives the unnormalized $2\times2$ density
\begin{equation}
\label{eq:spin-retained-density}
(\rho_\Lambda)_{s_\Lambda s'_\Lambda}
=\sum_{s_b,s'_b,\eta}(\rho_b)_{s_b s'_b}
\mathcal M_{s_b s_\Lambda\eta}
\mathcal M_{s'_b s'_\Lambda\eta}^{*}.
\end{equation}
An equivalent matrix form makes the dependence on parent spin preparation explicit. 
For each fixed unobserved nucleon-spin configuration $\eta$, collect the coherent amplitudes into a matrix $\mathsf A_\eta$ with daughter-spin rows and parent-spin columns. The daughter density is then obtained by applying these amplitude maps to $\rho_b$ and summing over $\eta$,

\begin{equation}
\label{eq:spin-amplitude-map}
[\mathsf A_\eta]_{s_\Lambda s_b}=\mathcal M_{s_b s_\Lambda\eta},
\qquad
\rho_\Lambda=\sum_\eta\mathsf A_\eta\rho_b\mathsf A_\eta^\dagger,
\qquad
\operatorname{Tr}\rho_\Lambda\big|_{\rho_b=\mathbf1/2}
=H^\dagger K H.
\end{equation}
For any physical parent density $\rho_b$, this representation gives a Hermitian, positive semidefinite daughter density whose trace gives the decay intensity for the specified parent spin density. 
The unpolarized choice recovers the scalar intensity in Eq.~(\ref{eq:unified-gram-master}). 
The indexed form and the matrix form are compared in Appendix B.

\FloatBarrier
\subsection{Physical Pauli responses and spin-inserted kernels}
\label{sec:3-4}

The $\Lambda$ polarization is resolved along axes defined by the three-body decay momenta. 
In the $\Lambda_b^0$ rest frame, the longitudinal direction $L$ follows the $\Lambda$ momentum, and the transverse direction $T$ follows the projection of the proton momentum perpendicular to the $\Lambda$ flight direction. 
The normal direction $N$ completes the right-handed triad, with Cartesian labels $(1,2,3)=(T,N,L)$. 
These event-dependent directions are related to the spin axes in the $\Lambda$ rest frame by the adopted canonical, rotation-free boost. 
Their explicit definitions are given in Section~\ref{sec:4-2}.

At each phase-space point $\Phi$, the unitary transformation $U_\Lambda(\Phi)$ maps the common daughter-spin components to the event-dependent measurement basis. 
The daughter-spin operators in the common basis and the corresponding conversion of the parent density are given by
\begin{equation}
\label{eq:spin-physical-operators}
\Sigma_0^\Lambda=\mathbf1,
\qquad
\Sigma_i^\Lambda(\Phi)=U_\Lambda^\dagger(\Phi)\sigma_iU_\Lambda(\Phi),
\qquad
\rho_b^{\mathrm{common}}(\Phi)
=U_b^\dagger(\Phi)\rho_b^{\mathrm{reference}}U_b(\Phi),
\qquad i=1,2,3.
\end{equation}
Through $U_\Lambda(\Phi)$, the operators $\Sigma_i^\Lambda(\Phi)$ represent the chosen daughter-spin directions in the common basis. 
Because the measurement axes depend on the decay momenta, the corresponding spin operators also depend on the phase-space point.
The symbol $\rho_b$ below denotes this common-basis matrix.
The daughter density contains the information needed for polarization. 
Its trace gives the decay intensity $W_0$, while its traces with the spin operators give the three spin responses $W_i$.
Their ratios give the pointwise polarization in the specified measurement frame where $W_0>0$,
\begin{equation}
\label{eq:spin-pauli-responses}
W_\mu=\operatorname{Tr}(\rho_\Lambda\Sigma_\mu^\Lambda),
\qquad \mu=0,1,2,3,
\qquad
P_{\Lambda i}=\frac{W_i}{W_0},
\qquad
\rho_\Lambda=\frac12\left(W_0\mathbf1+\sum_{i=1}^3W_i\Sigma_i^\Lambda\right).
\end{equation}
Hermiticity of $\rho_\Lambda$ makes the Pauli responses $W_\mu$ real. Positivity implies $W_0\geq0$ and $\sqrt{W_1^2+W_2^2+W_3^2}\leq W_0$. 
The spin responses $W_i$ are signed, and their ratios to $W_0$ define the pointwise polarization where $W_0>0$.
The longitudinal, transverse, and normal components are $W_L=W_3$, $W_T=W_1$, and $W_N=W_2$, with the same relabeling for the represented operators and kernels.

These physical Pauli responses can also be evaluated as bilinears in the same dynamical coefficients used for the scalar intensity. Inserting the parent density and represented daughter-spin operator into the common-spin matrix defines the coefficient-space kernels $K_\mu[\rho_b]$ through
\begin{equation}
\label{eq:spin-inserted-kernels}
K_\mu[\rho_b]
=X^\dagger\left(\rho_b^T\otimes\Sigma_\mu^\Lambda\otimes\mathbf1_\eta\right)X,
\qquad
W_\mu=H^\dagger K_\mu[\rho_b]H,
\qquad
K_0[\mathbf1/2]=\frac12X^\dagger X=K.
\end{equation}
For the existing row ordering $h=(s_b,s_\Lambda,\eta)$, the tensor factors act on the parent-spin space, the daughter-spin space, and the proton and antiproton spin spaces, respectively. 
The Hermitian kernels $K_\mu[\rho_b]$ act on the coefficient space of the current-product representation, which is distinct from the physical daughter-spin space of $\rho_\Lambda$. 
The scalar kernel $K_0[\rho_b]$ is positive semidefinite, whereas the signed spin kernels $K_i[\rho_b]$ need not be. 
The transpose of $\rho_b$ follows from the row-index convention used for $X$. 
The indexed equivalence is given in Appendix B.

\FloatBarrier
\section{Kinematics, angular observables and polarization}
\label{sec:4}

The scalar intensity and daughter-spin responses of Section~\ref{sec:3} become observables only after the integration region, angular weights and physical spin axes are specified. Section~\ref{sec:4-1} relates the Dalitz invariants to pair-frame angles and mass projections, while Section~\ref{sec:4-2} supplies the daughter-spin frames and oriented phase-space measure. With this geometry fixed, Section~\ref{sec:4-3} integrates the scalar and spin responses into weighted numerators and partial rates over common regions. 
Their normalized ratios give the hemisphere asymmetries and Legendre moments of Section~\ref{sec:4-4} and the $\Lambda$ polarization of Section~\ref{sec:4-5}. 
Finally, Section~\ref{sec:4-6} constructs the conjugate-rate combinations required for direct CP asymmetries and CP-averaged branching fractions.

\FloatBarrier
\subsection{Dalitz kinematics and invariant-mass projections}
\label{sec:4-1}

For an unpolarized parent with all final spins summed, a common rotation of the three momenta leaves the intensity unchanged. 
Two independent pair invariants specify the internal momentum configuration, while the pair-frame angles reparameterize the same Dalitz density rather than introducing further internal coordinates~\cite{MikhasenkoEtAl2020}. 
The pair masses $m(Y\bar p)$ and $m(p\bar p)$ are useful for examining where the individual contributions of the adopted model are concentrated and where they overlap. 
The Dalitz coordinates are $s_{13}=m^2(Y\bar p)$ and $s_{23}=m^2(p\bar p)$, with $s_{12}+s_{13}+s_{23}=M^2+m_Y^2+2m_p^2$. 
With the parent-spin average included, the differential decay density $d^2\Gamma_Y/\allowbreak (ds_{13}\,\allowbreak ds_{23})$ and branching fraction $\mathcal B_Y$ are~\cite{PDG2025}
\begin{equation}
\label{eq:unified-dalitz-rate}
\frac{d^2\Gamma_Y}{ds_{13}\,ds_{23}}
=\frac{\overline{|\mathcal M_Y|^2}}{256\pi^3M^3},
\qquad
\mathcal B_Y={\tau_{\Lambda_b}}\Gamma_Y.
\end{equation}
Integrating the Dalitz density over the same phase-space region gives the branching fraction in Eq.~(\ref{eq:integrated-overlap}), where the contributions of the two assignments and their interference are displayed explicitly.
For a pair $(i,\allowbreak j)$ of invariant $s$, denote the spectator mass by $m_s$. Let $z$ be the cosine of the angle between particle $i$ and the opposite spectator direction in the pair rest frame. The invariant of particle $j$ and the spectator is
\begin{equation}
\label{eq:unified-angle-map}
\begin{aligned}
s_{\mathrm{other}}&=A(s)-J(s)z,\\
A(s)&=m_s^2+m_j^2+
\frac{(M^2-m_s^2-s)(s+m_j^2-m_i^2)}{2s},\\
J(s)&=\frac{\sqrt{\lambda(M^2,m_s^2,s)\lambda(s,m_i^2,m_j^2)}}{2s},
\end{aligned}
\end{equation}
where the Källén function is $\lambda(x,\allowbreak y,\allowbreak z)=x^2+y^2+z^2-2xy-2xz-2yz$. 
The exact physical region obeys
\begin{equation}
\label{eq:unified-dalitz-boundaries}
(m_i+m_j)^2\le s\le(M-m_s)^2,
\qquad
A(s)-J(s)\le s_{\mathrm{other}}\le A(s)+J(s).
\end{equation}
The Jacobian $J(s)$ relates the pair-angle variable $z$ to the second squared invariant at fixed $s$.
In pair-invariant-angle coordinates $(s,\allowbreak z)$, the differential decay density is $J(s)\overline{|\mathcal M|^2}/\allowbreak (256\pi^3M^3)$.
In squared-invariant coordinates, the Dalitz branching densities $D_A$ instead satisfy
\begin{equation}
\label{eq:dalitz-branching-densities}
\begin{aligned}
D_A(s_{13},s_{23})
&=\frac{d^2\mathcal B_A}{ds_{13}\,ds_{23}}
=\frac{\tau_{\Lambda_b}}{256\pi^3M^3}
W_A(s_{13},s_{23}),\\
D_{\mathrm{inc}}&=D_{13}+D_{23},\qquad
D_{\mathrm{coh}}=D_{\mathrm{inc}}+D_{\mathrm{int}}.
\end{aligned}
\end{equation}
Here $W_A$ includes the weak normalization of the spin-averaged contribution. 
Integrating $D_A$ over one allowed squared invariant gives the projection in the other. Converting to $m_{ij}=\sqrt{s_{ij}}$ then gives $d\mathcal B/\allowbreak dm_{ij}=2m_{ij}\,\allowbreak d\mathcal B/\allowbreak ds_{ij}$.

The same kinematics determines how the pair thresholds appear in the Dalitz plot.
At the $Y\bar p$ threshold, $J(s_{13})=0$ and the allowed $s_{23}$ interval collapses to $A(s_{13})$, above the $p\bar p$ threshold. Conversely, low $s_{23}$ selects a boundary region with larger $s_{13}$. The two pair thresholds are therefore kinematically separated. At either exact pair threshold, the angle-to-invariant Jacobian vanishes. Section~\ref{sec:5-3} combines this geometry with the timelike inputs to interpret the assignment densities and mass projections.

\FloatBarrier
\subsection{Spin frames and oriented phase space}
\label{sec:4-2}

To define the $\Lambda$ polarization, spin directions must be specified for each decay configuration. 
The decay momenta first define an event frame in the parent rest frame. 
The corresponding axes are then related to the $\Lambda$ rest frame, and the oriented phase-space measure is introduced for integration.
For $\Lambda_b^0(P)\to\Lambda(p_1)p(p_2)\bar p(p_3)$, all three-momenta below are evaluated in the parent rest frame.
The daughter-flight direction is $\hat{\mathbf z}=\mathbf p_1/|\mathbf p_1|$. 
The momentum $p_2$ of the proton produced directly in the $\Lambda_b^0$ decay defines the visible decay plane.
The right-handed longitudinal, in-plane transverse and normal axes are then
\begin{equation}
\label{eq:spin-visible-plane-frame}
\hat{\mathbf e}_L=\hat{\mathbf z}_{pp}=\hat{\mathbf z},\qquad
\hat{\mathbf e}_N=\hat{\mathbf y}_{pp}
=\frac{\hat{\mathbf z}\times\mathbf p_2}
{|\hat{\mathbf z}\times\mathbf p_2|},\qquad
\hat{\mathbf e}_T=\hat{\mathbf x}_{pp}
=\hat{\mathbf y}_{pp}\times\hat{\mathbf z}.
\end{equation}
The transverse axis $\hat{\mathbf e}_T$ follows the proton's transverse momentum projection.
The spin responses $W_L,\allowbreak W_T,\allowbreak W_N$ retain daughter-spin information absent from the scalar density and depend on the same two internal invariants.
The corresponding frame is undefined at collinearity, $\hat{\mathbf z}\times\mathbf p_2=0$, or zero daughter momentum. These zero-measure boundaries are treated by a continuous limit or excluded explicitly.

The daughter-rest-frame axes retain the orientation of the parent-frame triad under the canonical, rotation-free boost along the daughter velocity~\cite{Marangotto2020}. For $E_\Lambda=p_1^0$ and $p_\Lambda=|\mathbf p_1|$, the boost maps each rest spin vector $(0,\allowbreak \hat{\mathbf e}_i)$ to
\begin{equation}
\label{eq:spin-canonical-axis-four-vectors}
a_L^\mu=\left(\frac{p_\Lambda}{m_\Lambda},
\frac{E_\Lambda}{m_\Lambda}\hat{\mathbf z}\right),\qquad
a_T^\mu=(0,\hat{\mathbf e}_T),\qquad
a_N^\mu=(0,\hat{\mathbf e}_N),\qquad
a_i\cdot p_1=0,\qquad a_i\cdot a_j=-\delta_{ij}.
\end{equation}
If $U_\Lambda(\Phi)$ maps the common external-spin basis to this physical frame, then $\rho_\Lambda^{pp}=U_\Lambda\rho_\Lambda U_\Lambda^\dagger$ and $\Sigma_i^\Lambda=U_\Lambda^\dagger\sigma_iU_\Lambda$. A helicity construction or sequence of noncollinear boosts must include its corresponding spin rotation in $U_\Lambda$ to preserve the physical insertion. Rotating the density and its spin operators together leaves their contraction invariant.

To integrate these spin responses, the phase-space measure includes the directions that specify the parent spin and the event-dependent daughter-spin axes.
For the pair invariant $s=s_{23}$, these directions are described by the $\Lambda$ solid-angle measure $d\Omega_\Lambda$ and the proton solid-angle measure $d\Omega_{\mathrm{pair}}=dz_{23}\,d\varphi_{\mathrm{pair}}$ in the proton-antiproton rest frame.
The polarization components are defined with respect to the event-plane axes introduced above. 
For the integration coordinates, an auxiliary right-handed triad $(\hat{\mathbf x}_0,\hat{\mathbf y}_0,\hat{\mathbf z})$ is chosen independently of the proton direction. 
The pair-frame axes are $\hat{\mathbf z}_q=-\hat{\mathbf z}$, $\hat{\mathbf y}_q=-\hat{\mathbf y}_0$, and $\hat{\mathbf x}_q=\hat{\mathbf x}_0$. 
This choice is consistent with the definition of $z_{23}$ in Section~4.1, measured relative to the direction opposite to the spectator $\Lambda$.

Let $\Delta_{\mathrm K}(u,\allowbreak v,\allowbreak w)=\lambda(u,\allowbreak v,\allowbreak w)$ denote the Källén function of Section~4.1, and set $\Delta_M(s)=\Delta_{\mathrm K}(M^2,\allowbreak m_\Lambda^2,\allowbreak s)$ and $\Delta_p(s)=\Delta_{\mathrm K}(s,\allowbreak m_p^2,\allowbreak m_p^2)$. 
With the relativistic phase-space convention underlying Eq.~(\ref{eq:unified-dalitz-rate}), the oriented measure is
\begin{equation}
\label{eq:spin-oriented-phase-space}
\frac{d\Phi_3}{2M}
=\mathcal C(s)\,ds\,d\Omega_\Lambda\,d\Omega_{\mathrm{pair}},\qquad
\mathcal C(s)=\frac{\sqrt{\Delta_M(s)\Delta_p(s)}}{4096\pi^5M^3s},\qquad
4m_p^2\le s\le(M-m_\Lambda)^2.
\end{equation}
The scalar pair-chart measure in Section~\ref{sec:4-1} integrates over the global orientation and the pair azimuth. 
To recover that measure without counting their volume twice, the oriented measure can be expressed using normalized angular measures,
\begin{equation}
\label{eq:spin-oriented-scalar-recovery}
\frac{d\Phi_3}{2M}
=\frac{J(s)}{256\pi^3M^3}\,ds\,dz_{23}\,
\frac{d\Omega_\Lambda}{4\pi}\,
\frac{d\varphi_{\mathrm{pair}}}{2\pi},\qquad
J(s)=\frac{\sqrt{\Delta_M(s)\Delta_p(s)}}{2s},\qquad
8\pi^2\mathcal C(s)=\frac{J(s)}{256\pi^3M^3}.
\end{equation}
Integrating the normalized orientation measures recovers the scalar measure of Eq.~(\ref{eq:unified-dalitz-rate}). 

\FloatBarrier
\subsection{General weighted, regional and binned observables}
\label{sec:4-3}

Weighted observables select angular or spin information from the decay distribution in a specified phase-space region. 
The weight determines the quantity being measured, while the chosen frame and axes define its angular or spin convention~\cite{ChenGeng2002}.

To express several integrated observables using common hadronic responses and weak coefficients, the decomposition $H=(G_F/\sqrt 2)Dg$ separates the phase-space-dependent hadronic functions in $D(\Phi)$ from the weak-current combinations in $g$.
The coefficient vector $g$ also includes the constant additional relative strong phase $\delta_Y$.
The remaining phase-space dependence and the chosen weight are collected in the integrated response matrix $R[w]$, whose contraction with $g$ gives the corresponding rate or weighted numerator,
\begin{equation}
\label{eq:unified-response}
R[w]={\tau_{\Lambda_b}}\frac{G_F^2}{2}
\int\frac{d\Phi_3}{2M}\,
w(\Phi)D^\dagger(\Phi)K(\Phi)D(\Phi).
\end{equation}
With the hadronic inputs and weak coefficients fixed and the additional phase taken to be constant, $g$ is independent of phase space.
The branching fraction $\mathcal B$ and normalized weighted observable $\langle w\rangle$ are

\begin{equation}
\label{eq:unified-weighted-observable}
\mathcal B=g^\dagger R[1]g,\qquad
\langle w\rangle=\frac{g^\dagger R[w]g}{g^\dagger R[1]g}.
\end{equation}
To describe a selected phase-space region, introduce the indicator $\chi_\Omega(\Phi)$, which equals one for $\Phi\in\Omega$ and zero otherwise.
The partial branching fraction is $g^\dagger R[\chi_\Omega]g$, and dividing it by $\mathcal B=g^\dagger R[1]g$ gives the selected event fraction.
This construction also applies to the proton-antiproton mass selection used in the $\Lambda$ measurement.
A signed weight can give a negative weighted numerator even though the decay density is nonnegative. 
For real $w$, the matrix $R[w]$ is Hermitian but need not be positive semidefinite.

To apply the same integrated-response construction to polarization, replace $K$ by the spin-inserted kernel $K_\mu[\rho_b]$ of Section~\ref{sec:3-4}.
Use the normalized parent density $\rho_b$ and physical daughter axes consistently at each phase-space point $\Phi$, with the same hadronic map $D$ and coefficient vector $g$, to obtain
\begin{equation}
\label{eq:spin-integrated-response}
R_\mu[w;\rho_b]
={\tau_{\Lambda_b}}\frac{G_F^2}{2}
\int\frac{d\Phi_3}{2M}\,w(\Phi)
D^\dagger(\Phi)K_\mu[\rho_b](\Phi)D(\Phi),
\qquad R_0[w;\mathbf1/2]=R[w].
\end{equation}
For real weights, the responses $R_\mu[w;\rho_b]$ are Hermitian. Positivity is guaranteed for the scalar response when the weight is nonnegative.
The weight selects or emphasizes regions of phase space in the integrated spin response.
With positive scalar normalization, the regional spin moment $\mathcal P_i[w;\Omega]$ is
\begin{equation}
\label{eq:spin-weighted-moment}
\mathcal P_i[w;\Omega]
=\frac{g^\dagger R_i[w\chi_\Omega;\rho_b]g}
{g^\dagger R_0[\chi_\Omega;\rho_b]g},
\qquad i=L,T,N.
\end{equation} 
Here $i=L,\allowbreak T,\allowbreak N$ labels a component along the physical daughter-rest-frame axes. 
The event-dependent spin operator in Eq.~(\ref{eq:spin-physical-operators}) expresses those axes in the common spin basis used for the amplitudes.

The effect of interference on a normalized angular or spin observable depends on both its numerator and its rate denominator.
The coherent and incoherent quantities are evaluated with the same region, axes, parent density and inputs. 
For $a=\mathrm{coh},\mathrm{inc}$, write $\mathcal O_a=\mathcal N_a/\mathcal B_a$, where $\mathcal N_a$ is the weighted numerator and $\mathcal B_a$ is the regional branching fraction. 
Define
$\mathcal N_{\mathrm{int}}
=\mathcal N_{\mathrm{coh}}-\mathcal N_{\mathrm{inc}}$
and
$\mathcal B_{\mathrm{int}}
=\mathcal B_{\mathrm{coh}}-\mathcal B_{\mathrm{inc}}$.
For positive coherent and incoherent rates,
\begin{equation}
\label{eq:normalized-interference-shift}
\mathcal O_{\mathrm{coh}}-\mathcal O_{\mathrm{inc}}
=
\frac{
\mathcal N_{\mathrm{int}}
-\mathcal O_{\mathrm{inc}}\mathcal B_{\mathrm{int}}
}{
\mathcal B_{\mathrm{coh}}
}.
\end{equation}
A positive interference contribution to the rate can therefore increase or decrease a normalized observable, depending on the interference numerator. 
It should be noticed that the interference term is not an independently physical component, so $\mathcal N_{\mathrm{int}}/\mathcal B_{\mathrm{int}}$ does not define its polarization, even when $\mathcal B_{\mathrm{int}}\ne0$.

\FloatBarrier
\subsection{Pair-angle distributions and moments}
\label{sec:4-4}

For an angular cosine $z$, the hemisphere asymmetry $A_{\mathrm{FB}}(\Omega)$ measures the net balance between the two angular halves. 
Complementary information is provided by the first and second Legendre moments $\langle P_1\rangle_\Omega$ and $\langle P_2\rangle_\Omega$, which resolve an odd angular component and an even anisotropy. 
Because these observables are normalized by the rate, they probe angular structure that is not specified by the total branching fraction alone~\cite{GratrexHopferZwicky2016}.
In a chosen region $\Omega$, they are
\begin{equation}
\label{eq:exact-angular-observables}
\begin{aligned}
A_{\mathrm{FB}}(\Omega)
&=\frac{\displaystyle\int_\Omega d\Phi_3\,
\operatorname{sgn}(z)\overline{|\mathcal M|^2}}
{\displaystyle\int_\Omega d\Phi_3\,\overline{|\mathcal M|^2}}, \quad
\langle P_\ell\rangle_\Omega
&=\frac{\displaystyle\int_\Omega d\Phi_3\,
P_\ell(z)\overline{|\mathcal M|^2}}
{\displaystyle\int_\Omega d\Phi_3\,\overline{|\mathcal M|^2}}.
\end{aligned}
\end{equation}
In the assignment-$13$ chart, $z_{13}$ is the cosine of the angle between $Y$ and the opposite spectator-proton direction. The cosine $z_{23}$ is defined using the pair proton and opposite spectator-$Y$ direction in the assignment-$23$ chart. 
Both angles are defined from the final-state momenta and are evaluated in their respective pair rest frames. 

Mass-dependent angular measurements in baryonic meson decays provide precedents for retaining pair mass alongside an angular projection~\cite{LHCbBaryonicAngular2014,CMSBaryonicMoments2019}. For a fixed chart with cosine $z$ and pair invariant $s$, let $\Omega_k$ be the intersection of the physical selected region with a declared bin in $s$. Using the branching measure $d\mathcal B=\tau_{\Lambda_b}\overline{|\mathcal M|^2}d\Phi_3/\allowbreak (2M)$, define
\begin{equation}
\label{eq:mass-binned-hemisphere}
\mathcal B_k=\int_{\Omega_k}d\mathcal B,\qquad
\mathcal N_k=\int_{\Omega_k}\operatorname{sgn}(z)\,d\mathcal B,\qquad
A_{\mathrm{FB},k}=\frac{\mathcal N_k}{\mathcal B_k},\qquad
\mathcal B_k>0.
\end{equation}
The coherent calculation includes the interference between the two assignment amplitudes, while the incoherent calculation adds their separate intensities without the cross term. 
Both use identical bins, axes, and inputs. 
Their differences give the interference contributions to the bin numerator and rate, which enter the normalized comparison in Eq.~\ref{eq:normalized-interference-shift}.
Summing the bin numerators and rates recovers the corresponding integrated asymmetry as $(\sum_k\mathcal N_k)/\allowbreak (\sum_k\mathcal B_k)$. With the chosen mass bins, the binwise numerator can distinguish small angular imbalances across the bins from compensating positive and negative regions.

Higher odd angular moments can contribute to the hemisphere asymmetry, so its directly integrated value need not equal the estimate based on the first Legendre moment. 
A mass selection can also restrict the angular range in the crossed pair frame. 
The selected angular distributions therefore depend on both the phase-space region and the amplitude within that region.

\FloatBarrier
\subsection{$\Lambda$ polarization}
\label{sec:4-5}

The $\Lambda$ polarization probes the spin structure of the decay and its dependence on the selected phase-space region. 
The daughter-spin density of Section~\ref{sec:3} is used to define the polarization components along the axes introduced in Section~\ref{sec:4-2}~\cite{ChenGengNg2002}.
For an unpolarized parent, $\rho_b=\mathbf1/ 2$, use the transported density $\rho_\Lambda^{pp}=U_\Lambda\rho_\Lambda U_\Lambda^\dagger$, with daughter-spin basis $(+,\allowbreak -)$ along $\hat{\mathbf e}_L$. The standard Pauli convention $\sigma_y=\bigl(\begin{smallmatrix}0&-i\\i&0\end{smallmatrix}\bigr)$ fixes

\begin{equation}
\label{eq:spin-visible-pauli-traces}
W_0=\rho_{++}^{pp}+\rho_{--}^{pp},\qquad
W_L=\rho_{++}^{pp}-\rho_{--}^{pp},\qquad
W_T=2\operatorname{Re}\rho_{+-}^{pp},\qquad
W_N=-2\operatorname{Im}\rho_{+-}^{pp}.
\end{equation}
Dividing each signed response by $W_0=\operatorname{Tr}\rho_\Lambda^{pp}$ gives
\begin{equation}
\label{eq:spin-visible-polarization-components}
P_L^{pp}
=\frac{\rho_{++}^{pp}-\rho_{--}^{pp}}{\operatorname{Tr}\rho_\Lambda^{pp}},\qquad
P_T^{pp}
=\frac{2\operatorname{Re}\rho_{+-}^{pp}}{\operatorname{Tr}\rho_\Lambda^{pp}},\qquad
P_N^{pp}
=\frac{-2\operatorname{Im}\rho_{+-}^{pp}}{\operatorname{Tr}\rho_\Lambda^{pp}}.
\end{equation}
The off-diagonal entry $\rho_{+-}^{pp}$ describes coherence between physical daughter-spin states.

Eq.~(\ref{eq:spin-weighted-moment}) has the following physical-response form for a region $\Omega$ with positive scalar rate and a real weight $w(\Phi)$
\begin{equation}
\label{eq:spin-regional-polarization}
\mathcal P_i[w;\Omega]
=\frac{\displaystyle\int_\Omega d\Phi_3\,wW_i}
{\displaystyle\int_\Omega d\Phi_3\,W_0},
\qquad i=L,T,N.
\end{equation}
At unit weight, $\mathcal P_i[1;\Omega]$ gives the selected rate-weighted polarization under the matched-region, transported-axis and integration conventions of Section~4.3.
The components probe different parts of the daughter density in the event plane. Interference shifts both the spin numerator and rate denominator according to Eq.~(\ref{eq:normalized-interference-shift}).



\FloatBarrier
\subsection{CP-conjugate rate observables}
\label{sec:4-6}

The direct CP asymmetry and CP-averaged branching fraction combine the particle and antiparticle rates in a selected phase-space region and are defined by
\begin{equation}
\label{eq:lambda-cp-observables}
\begin{aligned}
A_{\mathrm{CP}}^\Lambda(\Omega)
&=\frac{\mathcal B_\Lambda(\Omega)-\overline{\mathcal B}_\Lambda(\Omega)}
{\mathcal B_\Lambda(\Omega)+\overline{\mathcal B}_\Lambda(\Omega)},\quad
\mathcal B_{\Lambda,\mathrm{av}}(\Omega)
&=\frac{\mathcal B_\Lambda(\Omega)+\overline{\mathcal B}_\Lambda(\Omega)}{2}.
\end{aligned}
\end{equation}
Here $\mathcal B_\Lambda(\Omega)$ and $\overline{\mathcal B}_\Lambda(\Omega)$ refer to $\Lambda_b^0\to\Lambda p\bar p$ and $\bar\Lambda_b^0\to\bar\Lambda\bar p p$, respectively, in CP-related regions with the same invariant-mass cuts. 
The branching fractions use the common parent lifetime $\tau_{\Lambda_b}$.
The conjugate amplitude uses conjugated CKM factors and the antiparticle effective coefficients in Table~\ref{table:lambda-effective-coefficients}, with the same hadronic inputs~\cite{GengLiuTsai2021}.
The absorptive strong phases and the additional CP-even phase $\delta_\Lambda$ are retained.

\FloatBarrier
\section{Numerical results}
\label{sec:5}

The numerical analysis uses the observables of Section~\ref{sec:4} to determine how the adopted amplitudes distribute the two decay rates across momentum and spin configurations.  
Section~\ref{sec:5-1} specifies the reference inputs and the selected $\Lambda$ region.
Sections~\ref{sec:5-2} and~\ref{sec:5-3} establish the normalization and kinematic context by resolving the rate contributions and locating them in the Dalitz distributions and mass spectra. Against this background, Section~\ref{sec:5-4} examines the pair-angle distributions, moments and mass-resolved neutron cancellation, while Section~\ref{sec:5-5} determines how the $\Lambda$ selection and current interference alter the daughter polarization. 
Section~\ref{sec:5-6} discusses CP observables and the comparison with experiment.

\FloatBarrier
\subsection{Adopted inputs}
\label{sec:5-1}

The reference benchmark uses the pair-fit inputs~\cite{HuangHsiaoWangSun2022}, $N_c^{\mathrm{eff}}=3$, and zero additional relative phase, $\delta_Y=0$. 
Results cover the full Dalitz domain of Eq.~(\ref{eq:unified-dalitz-boundaries}) and the $\Lambda$ selection $m(p\bar p)<2.85\,\allowbreak \mathrm{GeV}$, equivalently $s_{23}<(2.85\,\allowbreak \mathrm{GeV})^2$.
The cut applies to both conjugate modes and restricts the crossed angular domain. 
The amplitudes $\mathcal M^{(q)}$ combine lattice-QCD transition form factors~\cite{DetmoldLehnerMeinel2015,DetmoldMeinel2016} with phenomenological timelike pair functions~\cite{HuangHsiaoWangSun2022}. 
The reference uses the nominal lattice fits, retaining all six covariant transition functions $f_i^N,\allowbreak g_i^N$ obtained from the helicity fits through Eq.~(\ref{eq:unified-lattice-conversion}). 
The nominal fits define the central predictions, while their higher-order alternatives are used in the partial uncertainty estimate. 
The neutron prescription is $f_i^n=f_i^p$, $g_i^n=g_i^p$ with positive relative isospin phase and physical masses. 
For the $\Lambda$ transition, the corresponding lattice form-factor fit is used~\cite{DetmoldLehnerMeinel2015,DetmoldMeinel2016}.
The densities $f_S^N,\allowbreak g_P^N$ follow Eq.~(\ref{eq:unified-transition-ward}). The transition conversion and released-fit conventions are specified in Sections A.2 and A.4.
The complete amplitudes use the operator contributions specified in Section~\ref{sec:2}.
The Wolfenstein truncations are $\lambda_u^{(d)}=A\lambda^3(\rho-i\eta)(1-\lambda^2/2)$, $\lambda_t^{(d)}=A\lambda^3(1-\rho+i\eta)$, $\lambda_u^{(s)}=A\lambda^4(\rho-i\eta)$ and $\lambda_t^{(s)}=-A\lambda^2$~\cite{GengLiuTsai2021}. 
The normalization constants follow the PDG Physical Constants
table~\cite{PDG2025}. The CKM benchmark and
$\overline{\mathrm{MS}}$ quark masses at
$\mu=2.5\,\mathrm{GeV}$ follow
Ref.~\cite{GengLiuTsai2021}, and the pair constants come from
the global fit of Ref.~\cite{HuangHsiaoWangSun2022}. The
particle masses and parent lifetime follow Ref.~\cite{PDG2026}.
The adopted numerical inputs are
\begin{equation}
\label{eq:central-input-values}
\begin{aligned}
&G_F=1.1663785(6)\times10^{-5}\,\mathrm{GeV}^{-2},
\\
&(\lambda,A,\rho,\eta)=(0.22650,0.790,0.141,0.357),\\
&(m_b,m_u,m_d,m_s)
=(4.88,0.00204,0.00442,0.0881)\,\mathrm{GeV},\\
&(C_\parallel,\delta C_\parallel,
C_{\overline\parallel},\delta C_{\overline\parallel})
=(150.8,31.9,27.4,-735.0)\,\mathrm{GeV}^4,\\
&(\bar C_\parallel,\delta\bar C_\parallel)
=(511.2,-317.8)\,\mathrm{GeV}^4,\\
&(C_D,C_F)=(-761.1,905.7)\,\mathrm{GeV}^6,\\
&M=5.61957(16)\,\mathrm{GeV},
\qquad
m_n=0.93956542194(48)\,\mathrm{GeV},\\
&m_\Lambda=1.115683(6)\,\mathrm{GeV},
\qquad
m_p=0.93827208943(29)\,\mathrm{GeV},\\
&\tau_{\Lambda_b}=1.465(9)\,\mathrm{ps}.
\end{aligned}
\end{equation}
The daughter mass is $m_Y=m_n$ in the neutron channel and $m_Y=m_\Lambda$ in the $\Lambda$ channel. 
The antiparallel vector-axial constant $C_{\overline\parallel}$ and scalar constant $\bar C_\parallel$ are distinct. 
With $L(t)=\ln(t/\allowbreak \Lambda_0^2)$, the pair form factors are~\cite{HuangHsiaoWangSun2022}
\begin{equation}
\label{eq:unified-updated-pair-shapes}
\begin{aligned}
(F_1^x,g_A^x,F_S^x,G_P^x)
=\frac{(C_{F_1}^x,C_{g_A}^x,C_{F_S}^x,C_{G_P}^x)}
{t^2L(t)^{2.148}},\quad
h_A^x=\frac{C_{h_A}^x}{t^3L(t)^{3.148}},
\quad
F_2^x=\frac{\mu_0^2}{tL(t)}F_1^x.
\end{aligned}
\end{equation}
The scales are $\Lambda_0=0.3\,\allowbreak \mathrm{GeV}$ and $\mu_0=1\,\allowbreak \mathrm{GeV}$. 
The Pauli form factor follows the adopted estimate in~\cite{GengHsiaoRodrigues2016,HuangHsiaoWangSun2022}.
The adopted neutral induced-axial prescription is $C_{h_A}^{dd}=C_F-C_D$, $C_{h_A}^{ss}=0$ and $C_{h_A}^{du}=C_{h_A}^{uu}-C_{h_A}^{dd}$. 

Uncertainties from the proton and $\Lambda$ transition form factors are estimated using the nominal and higher-order lattice fits~\cite{DetmoldLehnerMeinel2015,DetmoldMeinel2016}. 
The first uncertainty is obtained from the nominal transition-fit covariance, with shared fit parameters varied consistently across the relevant contributions and regions. 
For quantities marked by $\Delta$, the variance is evaluated from the linear response to the fitted parameters. 
The second is a partial higher-order estimate, given by the larger of the absolute central-value shift and the square root of the difference between the higher-order and nominal variances, with negative differences set to zero.
This joint prescription extends the separate lattice-fit treatments to both transition inputs. 
The unavailable covariance between the proton and $\Lambda$ fits is set to zero, and all other inputs are held fixed.

\FloatBarrier
\subsection{Decay rates and amplitude composition}
\label{sec:5-2}

The integrated branching fractions establish the relative sizes of the two channels and the contributions of their amplitudes.
The rates at the reference inputs are compared first, followed by their operator content and dependence on the effective-color parameter and relative phase.
At $N_c^{\mathrm{eff}}=3$ and $\delta_Y=0$, the coherent full-region rate is larger for the $\Lambda$ channel than for the neutron channel. 
With the nominal transition fits as the reference, the branching fractions and their partial transition uncertainties are
\begin{equation}
\label{eq:lambda-partial-uncertainties}
\begin{aligned}
\mathcal B_{n,\mathrm{full}}
&=(3.221\pm0.761_{\mathrm{tr,stat}}\pm2.056_{\mathrm{tr,HO}})\times10^{-6},\\
\mathcal B_{\Lambda,\mathrm{full}}
&=(7.000\pm0.875_{\mathrm{tr,stat}}\pm2.283_{\mathrm{tr,HO}})\times10^{-6},\\
\mathcal B_{\Lambda,\mathrm{cut}}
&=(3.021\pm0.335_{\mathrm{tr,stat}}\pm0.736_{\mathrm{tr,HO}})\times10^{-6}.
\end{aligned}
\end{equation}
The first errors arise from the nominal transition-fit covariance and the second are partial higher-order estimates, as described in Section~\ref{sec:5-1}.
Using the common parent mass and lifetime in Eq.~\eqref{eq:central-input-values} gives full widths $\Gamma_n=1.45\times10^{-18}\,\mathrm{GeV}$ and $\Gamma_\Lambda=3.14\times10^{-18}\,\mathrm{GeV}$.

\begin{table}[htbp]
\centering\small
\caption{
Branching contributions in units of $10^{-6}$ at $N_c^{\mathrm{eff}}=3$ and $\delta_Y=0$. Both modes use the reference inputs of Section~5.1. The selected $\Lambda$ row applies $m(p\bar p)<2.85\,\mathrm{GeV}$ and gives the particle rate. Interference is signed.
}
\label{table:two-mode-rates}
\renewcommand{\arraystretch}{1.16}
\setlength{\tabcolsep}{5pt}
\begin{fitwidthtable}
\begin{tabular}{@{}lrrrr@{}}
\toprule
\shortstack[l]{Mode\\and\\region} & \shortstack[r]{Assignment\\$13$} & \shortstack[r]{Assignment\\$23$} & Interference & \shortstack[l]{Coherent\\total} \tabularnewline
\midrule
Neutron, full & $3.29$ & $0.0742$ & $-0.140$ & $3.22$ \tabularnewline
Lambda, full & $3.84$ & $1.27$ & $1.90$ & $7.00$ \tabularnewline
Lambda, selected & $0.703$ & $1.16$ & $1.16$ & $3.02$ \tabularnewline
\bottomrule
\end{tabular}
\end{fitwidthtable}
\end{table}


Coherent overlap contributes to this channel contrast. In Table~\ref{table:two-mode-rates}, interference lowers the neutron incoherent sum $3.36\times10^{-6}$ by $4.17\%$ and raises the $\Lambda$ sum $5.10\times10^{-6}$ by $37.1\%$. Although $\mathcal M_{23}^{(d)}$ has a small isolated neutron rate, its cross term is linear in that amplitude with the other contribution held fixed and exceeds its isolated square in magnitude. The two assignments have more comparable strengths in the $\Lambda$ mode, allowing a larger relative interference contribution. Their overlap is constructive at the reference inputs, whereas the neutron overlap is destructive.
Selecting $m(p\bar p)<2.85\,\allowbreak \mathrm{GeV}$ retains $43.2\%$ of the $\Lambda$ particle rate, reverses the isolated-contribution hierarchy and raises the relative interference enhancement to $62.4\%$.

Table~\ref{table:operator-composition} separates the amplitudes into current-current terms $T$ and penguin terms $P=P_{\mathrm{QCD}}+P_{\mathrm{EW}}$. 
For the adopted weak inputs, $|\lambda_u^{(d)}/\lambda_t^{(d)}|$ is of order unity, while $|\lambda_u^{(s)}/\lambda_t^{(s)}| =\lambda^2\sqrt{\rho^2+\eta^2}\simeq0.0197$.
Together with the adopted coefficients and hadronic inputs, these weights give tree dominance in the neutron channel and penguin dominance in the $\Lambda$ channel.
Keeping only $T+P_{\mathrm{QCD}}$ gives branching fractions $3.21\times10^{-6}$ and $7.92\times10^{-6}$ in the neutron and $\Lambda$ channels, respectively. 
The electroweak terms shift these coherently to the full values in Table~\ref{table:operator-composition}. 
The vector-axial, scalar-pseudoscalar and cross contributions are $(2.87,0.396,-0.0443)\times10^{-6}$ for the neutron channel and $(2.30,3.08,1.62)\times10^{-6}$ for the $\Lambda$ channel. 
In the $\Lambda$ channel, the adopted neutral strange-current omission leaves the charged density term $\kappa_s^{SP}S_p^-D_{su}^+$ as the scalar-pseudoscalar
contribution. 
Its sizable square and positive interference with the vector-axial terms show the role of the charged-pair densities $F_S^{su}$ and $G_P^{su}$ in this prediction.

\begin{table}[htbp]
\centering\small
\caption{Full-region operator composition of the central amplitudes in units of $10^{-6}$. $T$ denotes current-current terms and $P$ the combined QCD and four-quark electroweak penguin terms. Each group includes both factorized assignments coherently. The signed $T$-$P$ overlap completes the full rate. Inputs and phases are those of Table~\ref{table:two-mode-rates}.}
\label{table:operator-composition}
\renewcommand{\arraystretch}{1.16}
\setlength{\tabcolsep}{5pt}
\begin{fitwidthtable}
\begin{tabular}{@{}lrrrr@{}}
\toprule
Mode & $\mathcal B_T$ & $\mathcal B_P$ & $\mathcal B_{TP}$ & \shortstack[l]{Full\\rate} \tabularnewline
\midrule
Neutron & $2.84$ & $0.408$ & $-0.0283$ & $3.22$ \tabularnewline
Lambda & $0.130$ & $7.04$ & $-0.170$ & $7.00$ \tabularnewline
\bottomrule
\end{tabular}
\end{fitwidthtable}
\end{table}
For the neutron channel, Ref.~\cite{GengHsiaoRodrigues2016} obtained a full-region central value of $2.0\times10^{-6}$ from the dominant tree contribution, compared with $3.22\times10^{-6}$ from the full amplitude used here. 
The earlier transition parameterization is
\begin{equation}
\label{eq:unified-set-a}
f_1^N(t)=g_1^N(t)=F_N(t),
\qquad
f_{2,3}^N(t)=g_{2,3}^N(t)=0.
\end{equation}
In a separate single-tree test, replacing this parameterization with the nominal lattice functions increases the rate by $61.53\%$. 
For the $\Lambda$ channel, Ref.~\cite{HsiaoGengRodrigues2019} reported a full-region central value of $3.2\times10^{-6}$, compared with the present prediction of $7.00\times10^{-6}$ for $\Lambda_b^0\to\Lambda p\bar p$. 
These comparisons are indicative because the calculations use different inputs and amplitude treatments, and the charge-averaging convention for the earlier $\Lambda$ result has not been established here.

\begin{figure}[tbp]
\centering
\includegraphics[width=\linewidth,height=0.5\textheight,keepaspectratio]{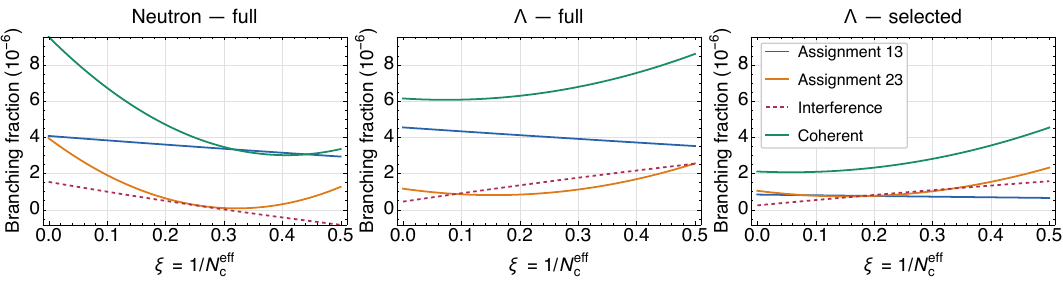}
\caption{Fixed-input effective-color sensitivity for the neutron full region, $\Lambda$ full region and $\Lambda$ region with $m(p\bar p)<2.85\,\allowbreak \mathrm{GeV}$. 
Curves show the isolated assignment contributions, signed interference and coherent total at reference transition and fitted pair-density inputs with relative phase $\delta=0$. 
The abscissa is the inverse effective color $\xi=1/ N_c^{\mathrm{eff}}$, with inverse effective color $\xi=0$ corresponding to the infinite-effective-color prescription. 
}
\label{fig:nc-comparison}
\end{figure}

The dependence on effective color illustrates how the relative current weights affect the rates and their interference.
Fig.~\ref{fig:nc-comparison} and Table~\ref{table:color-comparison} vary one real effective color parameter across the retained coefficient combinations, with $c_i^{\mathrm{eff}}$ and hadronic inputs fixed, over $\xi=1/ N_c^{\mathrm{eff}}\in[0, 1/ 2]$~\cite{GengHsiaoRodrigues2016}. 
At $\xi=0$, explicit inverse-color terms vanish at these fixed coefficients. 
The neutron response is nonmonotonic because the color-suppressed tree coefficient $a_2=-0.365+1.168/ N_c^{\mathrm{eff}}$ passes through zero near $N_c^{\mathrm{eff}}=3.2$. 
Surviving penguins keep $\mathcal M_{23}^{(d)}$ nonzero there, separating its isolated-rate minimum from the interference zero and from the coherent-rate minimum. 
Toward the infinite endpoint, this contribution becomes comparable with assignment $13$ and interference turns constructive.

\begin{table}[htbp]
\centering\small
\caption{Representative effective-color prescriptions within the continuous scan, with the nominal full lattice transition form factors $f_i^N,\allowbreak g_i^N$, fitted pair densities, all specified four-quark operators and relative phase $\delta=0$. Branching fractions are in units of $10^{-6}$. The $\Lambda$ selected column imposes $m(p\bar p)<2.85\,\allowbreak \mathrm{GeV}$. The infinite entry denotes inverse effective color $\xi=0$ at fixed effective coefficients.}
\label{table:color-comparison}
\renewcommand{\arraystretch}{1.16}
\setlength{\tabcolsep}{5pt}
\begin{fitwidthtable}
\begin{tabular}{@{}rrrr@{}}
\toprule
$N_c^{\mathrm{eff}}$ & Neutron, full & Lambda, full & Lambda, selected \tabularnewline
\midrule
$2$ & $3.34$ & $8.59$ & $4.53$ \tabularnewline
$3$ & $3.22$ & $7.00$ & $3.02$ \tabularnewline
$4$ & $3.98$ & $6.49$ & $2.53$ \tabularnewline
$\infty$ & $9.53$ & $6.14$ & $2.10$ \tabularnewline
\bottomrule
\end{tabular}
\end{fitwidthtable}
\end{table}

In the $\Lambda$ mode, interference remains constructive across the same domain. The full and selected rates decrease from two through four but reach shallow minima before the infinite endpoint, and the selected isolated contributions cross as their relative weights change.

To examine the relative-phase dependence of the interference, the strong phase is varied at $N_c^{\mathrm{eff}}=3$ while the isolated rates remain fixed. 
It is constant within each calculation and may differ between the two channels. 
The neutron and $\Lambda$ phases are denoted by $\delta_n$ and $\delta_\Lambda$, respectively,
\begin{equation}
\label{eq:comparative-phase-rates}
\begin{aligned}
\mathcal B_{n,\mathrm{full}}(\delta_n)
&=[3.360687-0.140111\cos\delta_n+0.303965\sin\delta_n]\times10^{-6},\\
\mathcal B_{\Lambda,\mathrm{full}}(\delta_{\Lambda})
&=[5.104369+1.895302\cos\delta_{\Lambda}+0.502314\sin\delta_{\Lambda}]\times10^{-6},\\
\mathcal B_{\Lambda,\mathrm{cut}}(\delta_{\Lambda})
&=[1.860557+1.160213\cos\delta_{\Lambda}+0.294235\sin\delta_{\Lambda}]\times10^{-6}.
\end{aligned}
\end{equation}
The complex overlaps generate both sine and cosine terms, so $\delta_{n/\Lambda}=0,\allowbreak \pi$ need not be the extrema. 
The full-region ranges are $[3.03,\allowbreak 3.70]\times10^{-6}$ for neutron and $[3.14,\allowbreak 7.07]\times10^{-6}$ for $\Lambda$.
The selected $\Lambda$ particle rate spans $[0.664,\allowbreak 3.06]\times10^{-6}$. 
The broader fractional $\Lambda$ response follows from its larger relative overlap. 

\FloatBarrier
\subsection{Dalitz structure and invariant-mass spectra}
\label{sec:5-3}

\begin{figure}[tbp]
\centering
\includegraphics[width=\linewidth,height=0.5\textheight,keepaspectratio]{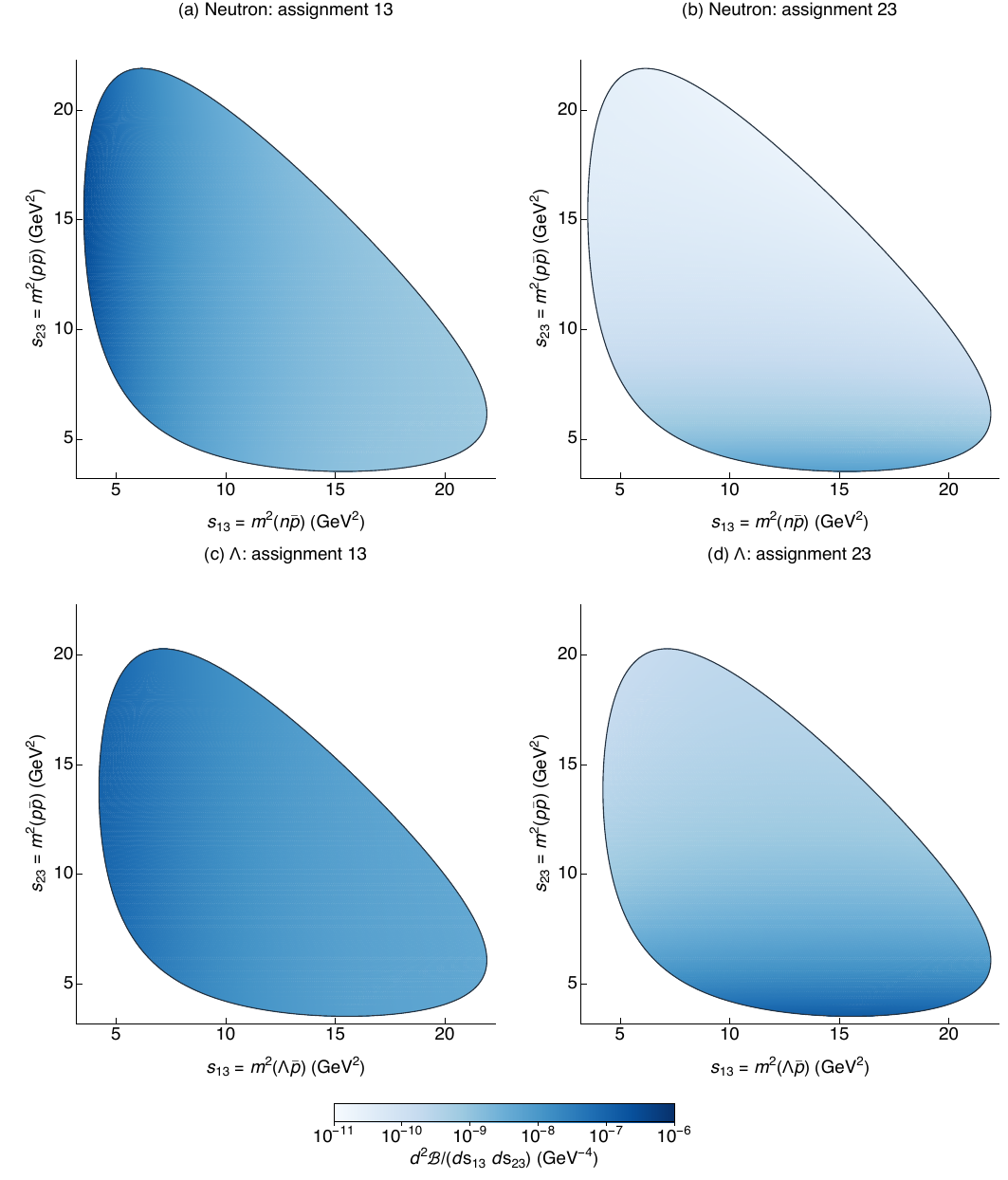}
\caption{Central absolute assignment densities for the neutron mode (top row) and $\Lambda$ mode (bottom row). The left panels show the assignment-$13$ Dalitz branching density $D_{13}$, and the right panels show the assignment-$23$ Dalitz branching density $D_{23}$. All panels use horizontal $s_{13}=m^2(Y\bar p)$, vertical $s_{23}=m^2(p\bar p)$ and one logarithmic density scale in $\mathrm{GeV}^{-4}$. 
}
\label{fig:dalitz-assignments}
\end{figure}

\begin{figure}[tbp]
\centering
\includegraphics[width=\linewidth,height=0.5\textheight,keepaspectratio]{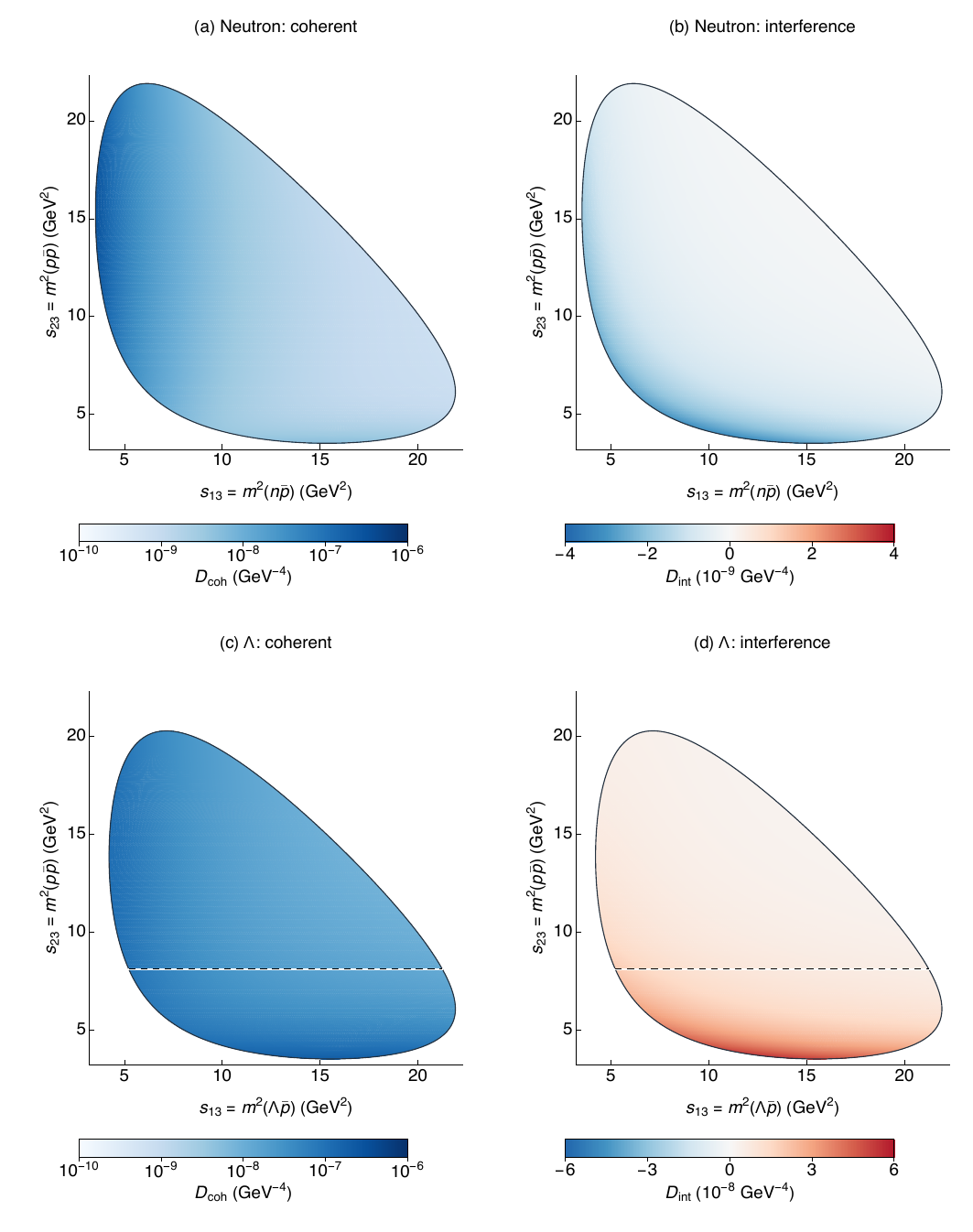}
\caption{Coherent Dalitz branching density $D_{\mathrm{coh}}$ (left) and signed assignment-interference density $D_{\mathrm{int}}$ (right) for the neutron mode (top) and $\Lambda$ mode (bottom), using the central full-amplitude inputs and relative phase $\delta=0$. The coherent panels share an absolute logarithmic scale in $\mathrm{GeV}^{-4}$. The interference panels use separately labeled symmetric linear scales, with negative values in blue and positive values in red. The dashed $\Lambda$ line marks $m(p\bar p)=2.85\,\allowbreak \mathrm{GeV}$.}
\label{fig:coherent-dalitz-comparison}
\end{figure}

The two current assignments contribute with different strengths across the Dalitz plot.
Their mass projections show how these regions and their interference contribute to the observed spectra.
The timelike pair functions in Eq.~(\ref{eq:unified-updated-pair-shapes}) favor low produced-pair masses. 
Fig.~\ref{fig:dalitz-assignments} shows that assignment $13$ is concentrated near small $s_{13}=m^2(Y\bar p)$, while assignment $23$ favors small $s_{23}=m^2(p\bar p)$.
Near either threshold band, the boundary geometry of Section~\ref{sec:4-1} places the crossed invariant in a higher interval. A broad high-mass crossed structure can consequently arise from low-mass pair production, without a dynamical preference for a heavy produced pair or a crossed invariant at its global maximum.
Ridge maxima and the interference sign also depend on transition functions, scalar terms and angular contractions. 
Comparing the assignment densities in Fig.~\ref{fig:dalitz-assignments} with the interference density in Fig.~\ref{fig:coherent-dalitz-comparison} resolves these effects.
Local peak height and integrated strength describe different properties of a density.
An assignment with a smaller total rate can still be more concentrated near a threshold.

Fig.~\ref{fig:coherent-dalitz-comparison} compares the coherent Dalitz densities with the signed interference densities. 
At the reference phase, the interference density is negative across the displayed neutron region and positive across the displayed $\Lambda$ region.
Its contribution to the total rate also depends on the absolute density. 
A large fractional change in a region with a small decay density can make only a small contribution to the integrated rate.
The coherent neutron distribution mainly follows the low-$s_{13}$ band of $\mathcal M_{13}^{(d)}$, whereas the $\Lambda$ distribution displays both bands and is most concentrated near the $p\bar p$ threshold.

\begin{figure}[tbp]
\centering
\includegraphics[width=\linewidth,height=0.5\textheight,keepaspectratio]{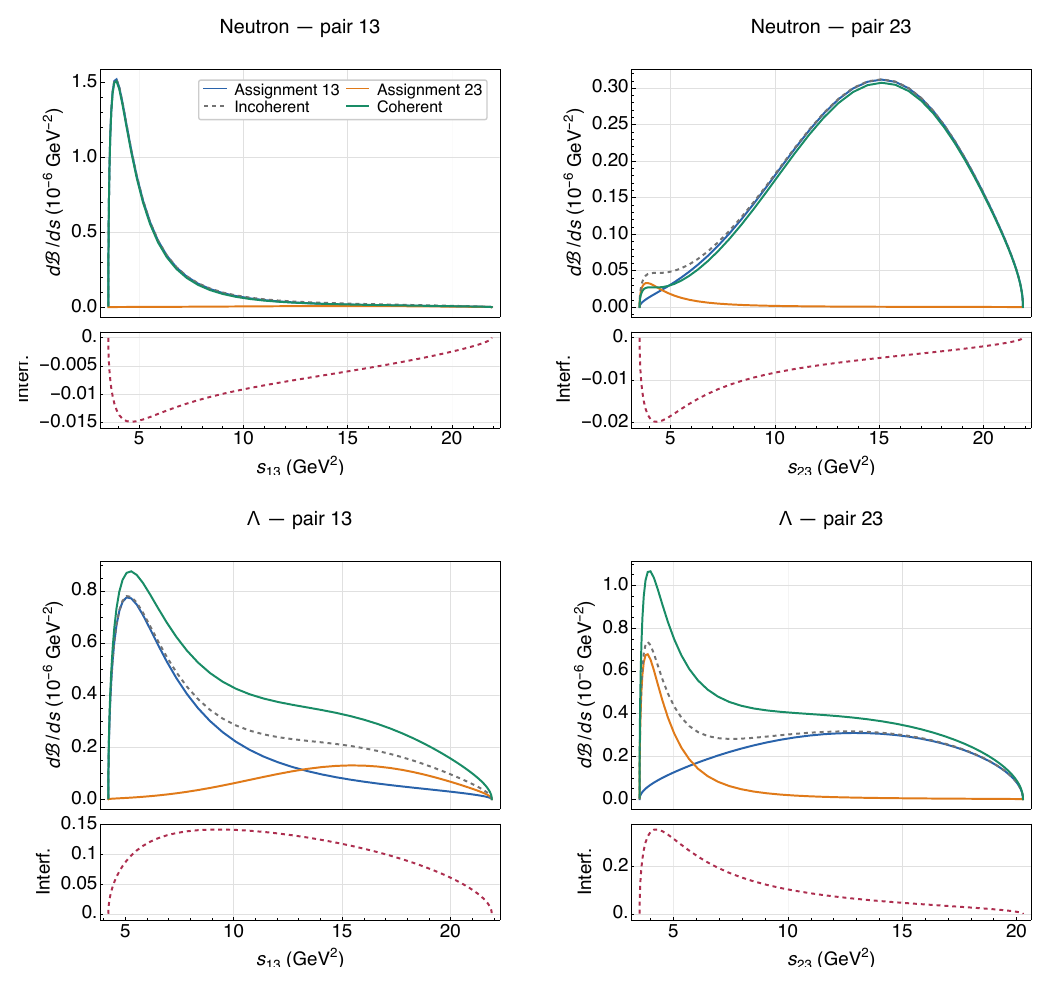}
\caption{Central assignment decomposition of the full-region mass-squared spectra for the neutron mode in the top row and $\Lambda$ mode in the bottom row. 
The main panels show assignment $13$, assignment $23$, their incoherent sum and their coherent total, while the lower strips show their signed differential interference $d\mathcal B_{\mathrm{int}}(\delta)/\allowbreak dx$. Columns correspond to $s_{13}$ and $s_{23}$. 
}
\label{fig:mass-comparison}
\end{figure}

Fig.~\ref{fig:mass-comparison} shows the mass projections of the two assignments, their coherent sum and their interference.
For a mass or angular variable $x$, the differential overlap $\mathcal I_x=d\mathcal I/\allowbreak dx$ of Eq.~(\ref{eq:integrated-overlap}) gives
\begin{equation}
\label{eq:differential-assignment-decomposition}
\begin{aligned}
\frac{d\mathcal B(\delta)}{dx}
&=\frac{d\mathcal B_{13}}{dx}
+\frac{d\mathcal B_{23}}{dx}
+2\operatorname{Re}\!\left(e^{i\delta}\mathcal I_x\right),\\
\frac{d\mathcal B_{\mathrm{int}}(\delta)}{dx}
&=\frac{d\mathcal B(\delta)}{dx}
-\frac{d\mathcal B_{\mathrm{inc}}}{dx},
\qquad
\frac{d\Gamma}{dx}=\frac{1}{\tau_{\Lambda_b}}\frac{d\mathcal B}{dx}.
\end{aligned}
\end{equation}
The absolute branching-fraction normalization preserves the additive current decomposition. 
The common lifetime conversion gives the corresponding differential widths.
The pair functions remain finite at threshold, while the phase-space Jacobian vanishes at the exact endpoints. 
Their combination produces smooth near-threshold spectral peaks. 
Both assignments contribute to both mass projections, so a peak in the observed $p\bar p$ spectrum alone cannot identify the current producing the pair.
In the neutron mode, $\mathcal M_{13}^{(d)}$ produces a sharp low-$s_{13}$ peak and a broad high-$s_{23}$ crossed projection. 
At the reference color choice, the smaller assignment-$23$ contribution leaves its low-$s_{23}$ enhancement subleading, while destructive interference suppresses both projections. 
The resulting contrast between the $n\bar p$ threshold enhancement and the broader $p\bar p$ spectrum is also present in Ref.~\cite{GengHsiaoRodrigues2016}.
The present two-current calculation shows how that pattern and its dependence on the color choice are affected by the additional contribution and its interference with the first.
For $\Lambda$, the larger $\mathcal M_{23}^{(s)}$ contribution produces a prominent low-$s_{23}$ peak, while $\mathcal M_{13}^{(s)}$ produces the low-$s_{13}$ peak. Each spectrum also receives the other assignment's crossed projection, and constructive interference adds a distinct mass dependence. The reference result thus retains the $\Lambda\bar p$ enhancement emphasized in the earlier $\Lambda_b^0$ study~\cite{HsiaoGengRodrigues2019} and also displays a substantial $p\bar p$ threshold band.
The cut $s_{23}<(2.85\,\allowbreak \mathrm{GeV})^2$ retains the low-$s_{23}$ enhancement preferentially over the broad high-$s_{23}$ projection of assignment $13$. 
Together with the larger relative constructive overlap, this produces the selected composition in Table~\ref{table:two-mode-rates}. The common Dalitz geometry also changes the crossed $Y\bar p$ spectrum, so both selected projections require integration over the cut region.

\begin{figure}[tbp]
\centering
\includegraphics[width=\linewidth,height=0.5\textheight,keepaspectratio]{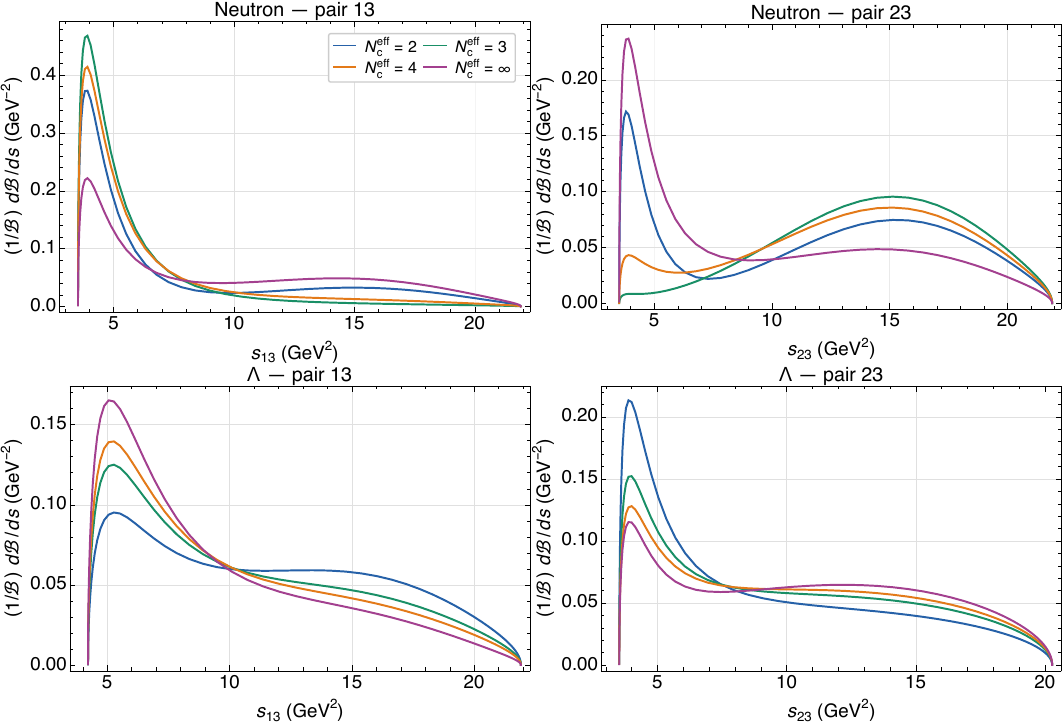}
\caption{Normalized full-region mass-squared spectra for the neutron mode (top row) and $\Lambda$ mode (bottom row) at $N_c^{\mathrm{eff}}=2,\allowbreak 3,\allowbreak 4,\allowbreak \infty$. 
Each curve is divided by its own coherent branching fraction, giving units $\mathrm{GeV}^{-2}$.}
\label{fig:color-mass-comparison}
\end{figure}

To examine how the relative phase changes the mass dependence, the isolated spectra are kept fixed while $\delta_Y$ is varied.
Changing $\delta_Y=0$ to $\pi$ reverses the differential interference, whereas $\delta_Y=\pi/2$ selects $-2\operatorname{Im}\mathcal I_x$.
The complex, projection-dependent overlap phase can therefore change the interference pattern with mass, consistently with the integrated response of Section~\ref{sec:5-2}.
Changing the effective color parameter alters the relative current weights and can therefore change the mass shapes as well as the total rates. Fig.~\ref{fig:color-mass-comparison} isolates this redistribution by dividing each curve by its own coherent branching fraction. In the neutron mode, the broad $s_{23}$ spectrum at three develops a prominent low-mass peak at the infinite endpoint, while the low-$s_{13}$ peak decreases and the enhanced assignment $23$ raises the crossed high-$s_{13}$ contribution. For $\Lambda$, among the displayed choices the low-$s_{23}$ peak is strongest at two and the low-$s_{13}$ peak at the infinite endpoint.
The changes in the two threshold regions therefore reflect the changing relative current weights at fixed hadronic inputs and relative phase.

\FloatBarrier
\subsection{Angular distributions and moments}
\label{sec:5-4}

\begin{figure}[tbp]
\centering
\includegraphics[width=\linewidth,height=0.5\textheight,keepaspectratio]{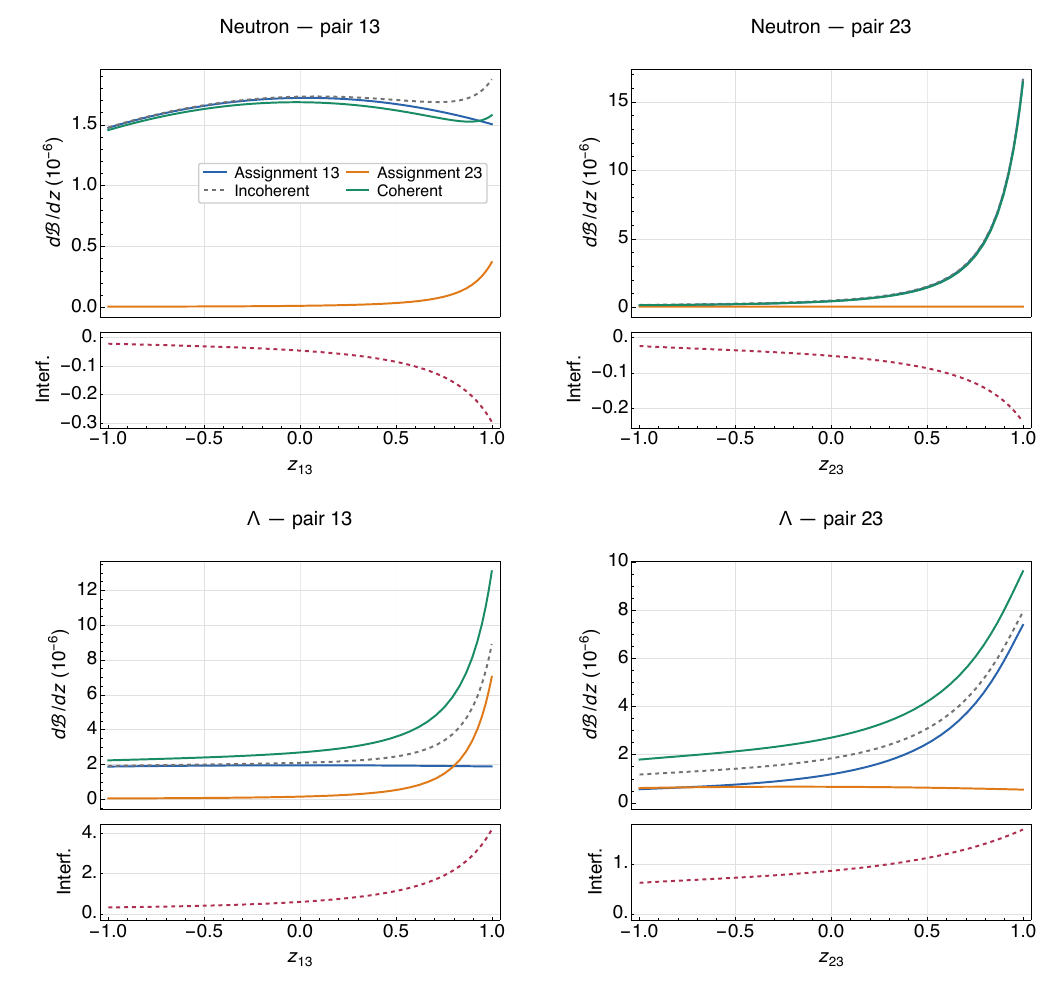}
\caption{Assignment decomposition of the full-region angular intensities for the neutron mode (top row) and $\Lambda$ mode (bottom row).
Columns correspond to $z_{13}$ and $z_{23}$, defined using the first pair baryon relative to the opposite spectator direction. Main curves retain absolute $d\mathcal B/\allowbreak dz$ normalization for the two assignments, their incoherent sum and their coherent total. Lower strips show signed differential interference $d\mathcal B_{\mathrm{int}}(\delta)/\allowbreak dz$ on separately labeled scales.}
\label{fig:angular-comparison}
\end{figure}

Figure~\ref{fig:angular-comparison} compares the angular spectra $d\mathcal B/\allowbreak dz_{13}$ and $d\mathcal B/\allowbreak dz_{23}$ for both decay channels, separating the two current assignments and their interference. At the reference inputs, assignment $13$ dominates both neutron projections, while the interference is negative. The $z_{23}$ spectrum has a much stronger forward enhancement than the $z_{13}$ spectrum. In the $\Lambda$ channel, assignment $23$ rises sharply toward the forward edge of the $z_{13}$ projection, whereas assignment $13$ produces the forward rise in the $z_{23}$ projection. Their interference is positive in both projections.

The two projections probe different angular dependences of each assignment. At fixed mass of its produced pair, a contribution is quadratic in the corresponding angular cosine within the adopted amplitude. In the crossed frame, the invariant mass of that pair varies with the angle as $s_{\mathrm{other}}=A(s)-J(s)z$, and so do its form factors. Their decrease with increasing pair mass can favor the forward region and generate higher angular moments even without interference. For example, in the neutron channel, assignment $13$ alone gives $A_{\mathrm{FB}}(z_{13})=0.004555$ in its own $n\bar p$ frame and $A_{\mathrm{FB}}(z_{23})=0.858$ in the crossed $p\bar p$ frame.

Higher odd moments can also contribute to the hemisphere asymmetry. Table~\ref{table:angular-comparison} compares the directly integrated asymmetries with the first two Legendre moments. For the full coherent neutron $z_{23}$ distribution, the estimate based on the first moment, $3\langle P_1\rangle/\allowbreak 2=1.04$, lies outside the physical interval, whereas direct integration gives $0.856$. For the two $\Lambda$ projections, the corresponding estimates, $0.364$ and $0.413$, also differ from the directly integrated values. The first moment alone therefore does not generally determine the hemisphere asymmetry.

\begin{table}[htbp]
\centering\small
\caption{Directly integrated full-region hemisphere asymmetries $A_{\mathrm{FB}}(\Omega)$ and normalized Legendre moments $\langle P_1\rangle_\Omega,\allowbreak \langle P_2\rangle_\Omega$ at effective color parameter $N_c^{\mathrm{eff}}=3$, relative phase $\delta=0$.
Selected $\Lambda$ Legendre moments $\langle P_1\rangle_\Omega,\allowbreak \langle P_2\rangle_\Omega$ are reported separately in Table~\ref{table:lambda-selected-angular}.}
\label{table:angular-comparison}
\renewcommand{\arraystretch}{1.16}
\setlength{\tabcolsep}{5pt}
\begin{fitwidthtable}
\begin{tabular}{@{}lrrrr@{}}
\toprule
\shortstack[l]{Mode\\and\\region} & Angle & $A_{\mathrm{FB}}$ & $\langle P_1\rangle$ & $\langle P_2\rangle$ \tabularnewline
\midrule
Neutron, full & $z_{13}$ & $0.000446$ & $0.001264$ & $-0.016977$ \tabularnewline
Neutron, full & $z_{23}$ & $0.856$ & $0.692$ & $0.437$ \tabularnewline
Lambda, full & $z_{13}$ & $0.311$ & $0.242$ & $0.133$ \tabularnewline
Lambda, full & $z_{23}$ & $0.380$ & $0.275$ & $0.107$ \tabularnewline
\bottomrule
\end{tabular}
\end{fitwidthtable}
\end{table}

\begin{figure}[tbp]
\centering
\includegraphics[width=\linewidth,height=0.5\textheight,keepaspectratio]{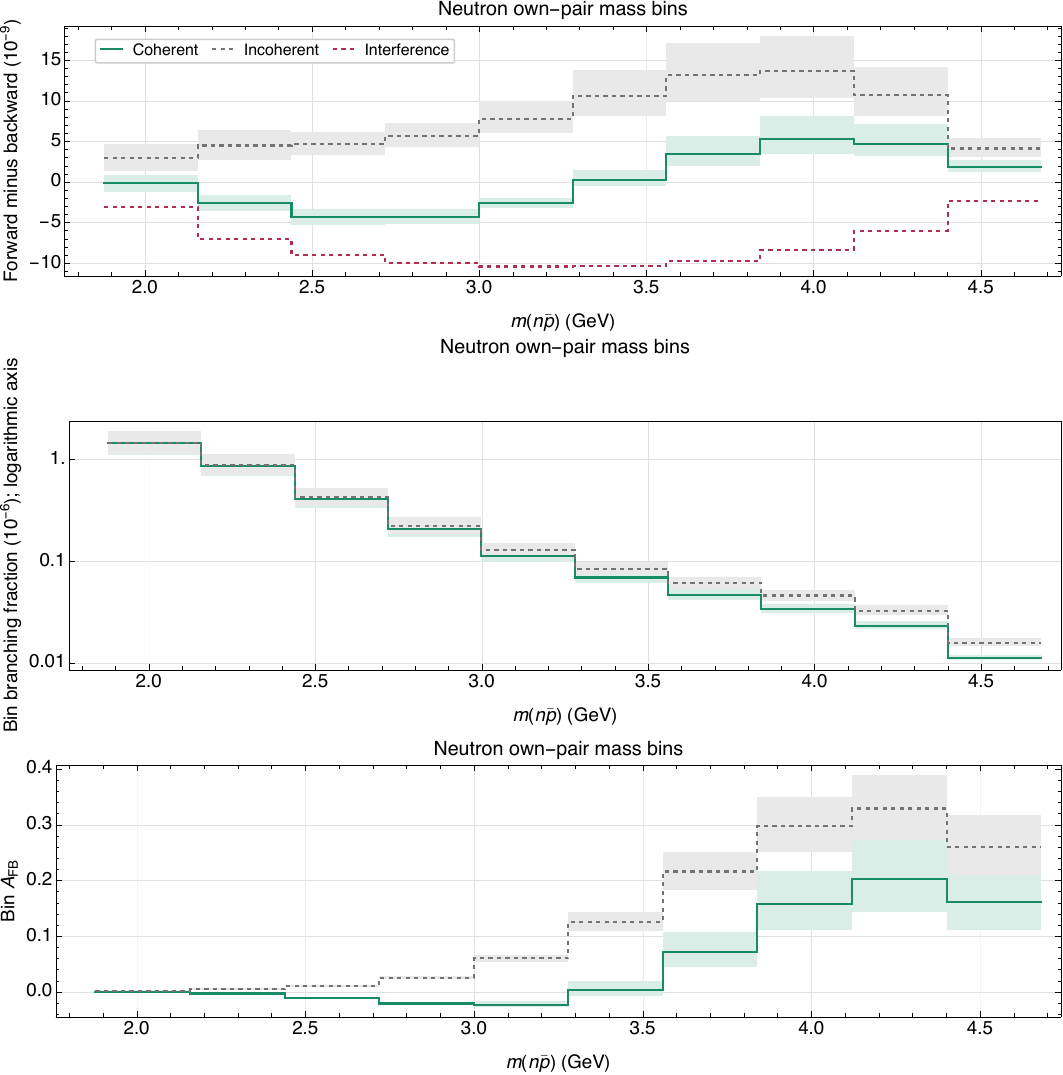}
\caption{Mass-resolved neutron forward-minus-backward branching contributions, bin rates and asymmetries at the reference inputs, $N_c^{\mathrm{eff}}=3$ and $\delta_n=0$. The ten equal-width $m(n\bar p)$ bins show the signed numerator $\mathcal N_k$ in units of $10^{-9}$, positive bin branching fraction $\mathcal B_k$ in units of $10^{-6}$ on a logarithmic scale, and bin asymmetry $A_{\mathrm{FB},k}$. Coherent and incoherent curves share the same bins and inputs. Signed interference is shown only in the numerator. Steps show the bin integrals and bin ratios of Eq.~(\ref{eq:mass-binned-hemisphere}). Shading gives nominal transition-only $16$th--$84$th percentile intervals from the shared proton-fit covariance.}
\label{fig:neutron-mass-angle}
\end{figure}

The small neutron $z_{13}$ asymmetry is sensitive to interference because its signed numerator can nearly cancel while the positive rate remains substantial. 
With both assignments included, the incoherent asymmetry is $0.023177$. 
Coherent interference reduces it to $0.000446$, although the total rate is lowered by only $4.17\%$. 
To show where this reduction occurs, Fig.~\ref{fig:neutron-mass-angle} resolves the forward-minus-backward branching contribution using Eq.~(\ref{eq:mass-binned-hemisphere}) in ten equal-width bins across the physical $m(n\bar p)$ interval, approximately $1.88$-$4.68\,\allowbreak \mathrm{GeV}$.
Within each displayed mass bin, the incoherent numerator is positive and the interference contribution is negative. 
Their sum gives negative coherent numerators in the first five bins and positive values in the last five. 
A second cancellation occurs when these coherent bin numerators are summed across mass.
The positive higher-mass contributions compensate the negative lower-mass contributions. 
The four highest-mass bins contain only $3.57\%$ of the coherent rate but supply nearly all of the positive numerator, and the bin asymmetry reaches about $0.20$ at $m(n\bar p)\simeq4.12$-$4.40\,\mathrm{GeV}$. 
The small integrated asymmetry is the residual of this regional compensation.
It is smaller than its partial transition-input variations and is specific to the reference inputs.
For $\Lambda$, constructive interference raises the $z_{13}$ asymmetry from the incoherent value $0.220$ to $0.311$ but lowers the $z_{23}$ asymmetry from $0.435$ to $0.380$. In Eq.~(\ref{eq:normalized-interference-shift}), the interference contribution to a weighted numerator competes with the increase of the common \mbox{rate denominator}. The angular dependence of the interference therefore determines whether a normalized asymmetry grows or decreases, even when the scalar interference is positive.

In the $\Lambda$ channel, the selection $m(p\bar p)<2.85\,\mathrm{GeV}$ modifies the two angular projections in different ways.
Table~\ref{table:lambda-selected-angular} shows a large selected $z_{13}$ asymmetry and a smaller $z_{23}$ asymmetry.
The geometry of the cut helps explain the former. 
For $J(s_{13})>0$, the requirement $m(p\bar p)<2.85\,\mathrm{GeV}$ gives $z_{13}>[A(s_{13})-(2.85\,\mathrm{GeV})^2]/J(s_{13})$ within the physical interval $[-1,1]$, preferentially removing the backward part of the crossed angular domain. 
The numerical asymmetry also depends on the amplitude distribution over the surviving region.

\begin{table}[htbp]
\centering\small
\caption{Directly integrated hemisphere asymmetries $A_{\mathrm{FB}}(\Omega)$ and normalized Legendre moments $\langle P_1\rangle_\Omega,\allowbreak \langle P_2\rangle_\Omega$ for $\Lambda$ with $m(p\bar p)<2.85\,\allowbreak \mathrm{GeV}$, at the same effective color parameter $N_c^{\mathrm{eff}}=3$, relative phase $\delta=0$ and full-amplitude inputs as Table~\ref{table:angular-comparison}.}
\label{table:lambda-selected-angular}
\renewcommand{\arraystretch}{1.16}
\setlength{\tabcolsep}{5pt}
\begin{fitwidthtable}
\begin{tabular}{@{}lrrrr@{}}
\toprule
\shortstack[l]{Mode\\and\\region} & Angle & $A_{\mathrm{FB}}$ & $\langle P_1\rangle$ & $\langle P_2\rangle$ \tabularnewline
\midrule
Lambda, selected & $z_{13}$ & $0.961$ & $0.746$ & $0.430$ \tabularnewline
Lambda, selected & $z_{23}$ & $0.185$ & $0.133$ & $0.0433$ \tabularnewline
\bottomrule
\end{tabular}
\end{fitwidthtable}
\end{table}

\begin{figure}[tbp]
\centering
\includegraphics[width=\linewidth,height=0.5\textheight,keepaspectratio]{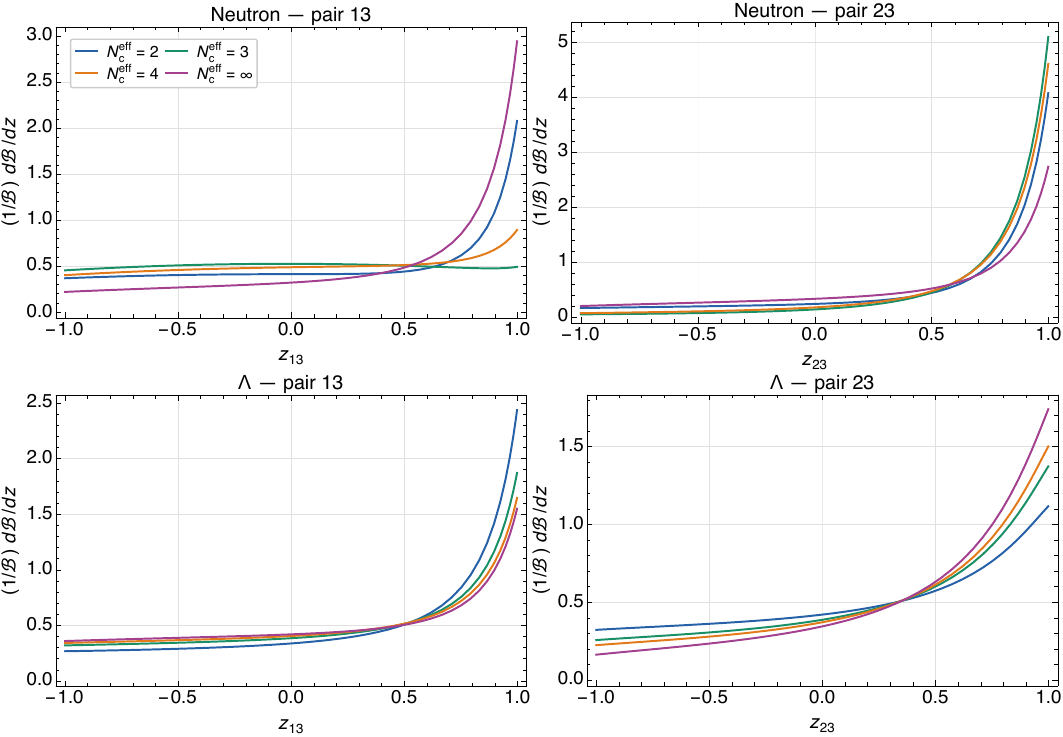}
\caption{Fixed-input effective-color sensitivity of the normalized full-region angular spectra for both modes, using the same four choices and hadronic inputs as the mass-shape comparison. Columns correspond to $z_{13}$ and $z_{23}$, and every curve uses its own coherent total.}
\label{fig:color-angular-comparison}
\end{figure}

The effective-color dependence of the mass spectra also affects the angular distributions. 
Fig.~\ref{fig:color-angular-comparison} and Table~\ref{table:color-asymmetries} show the corresponding normalized spectra and directly integrated asymmetries.
In the neutron channel, the $z_{13}$ asymmetry grows from its nearly cancelled reference value to $0.207$ at two and $0.468$ at the infinite endpoint.
At the latter endpoint, its two asymmetries become close, as do the isolated current strengths. 
For $\Lambda$, the two angular asymmetries move in opposite directions toward that endpoint. 
Considering both projections therefore resolves the directional redistribution associated with the assumed current weights beyond the information carried by a single total rate.

\begin{table}[htbp]
\centering\small
\caption{Directly integrated full-region forward-backward asymmetries $A_{\mathrm{FB}}(\Omega)$ of the coherent amplitudes at the representative effective-color choices, with reference transition and fitted pair inputs and relative phase $\delta=0$. Angles follow the same convention as Table~\ref{table:angular-comparison}. Every entry uses its own coherent rate denominator.}
\label{table:color-asymmetries}
\renewcommand{\arraystretch}{1.16}
\setlength{\tabcolsep}{5pt}
\begin{fitwidthtable}
\begin{tabular}{@{}rrrrr@{}}
\toprule
$N_c^{\mathrm{eff}}$ & Neutron $z_{13}$ & Neutron $z_{23}$ & Lambda $z_{13}$ & Lambda $z_{23}$ \tabularnewline
\midrule
$2$ & $0.207$ & $0.622$ & $0.417$ & $0.273$ \tabularnewline
$3$ & $0.000446$ & $0.856$ & $0.311$ & $0.380$ \tabularnewline
$4$ & $0.0970$ & $0.795$ & $0.266$ & $0.432$ \tabularnewline
$\infty$ & $0.468$ & $0.488$ & $0.227$ & $0.521$ \tabularnewline
\bottomrule
\end{tabular}
\end{fitwidthtable}
\end{table}

\FloatBarrier
\subsection{$\Lambda$ polarization}
\label{sec:5-5}

The $\Lambda$ polarization reveals how the two current contributions and their interference determine the daughter-spin distribution. The discussion first compares the contributions in the full and selected regions, then examines their mass dependence.
The rate-weighted regional moments $\mathcal P_i\equiv\mathcal P_i[1;\Omega]$ of Eq.~(\ref{eq:spin-regional-polarization}) are evaluated for an unpolarized parent, $\rho_b=\mathbf1/\allowbreak 2$, along the visible-plane axes of Section~4.2.
The longitudinal direction follows the $\Lambda$ momentum, the transverse direction points toward the proton from the $\Lambda_b^0$ decay, and the normal direction completes the oriented plane.
The regional polarization $\mathcal P_L$ measures the difference between the rate-weighted probabilities for the two longitudinal spin orientations. The components $\mathcal P_T$ and $\mathcal P_N$ probe the corresponding real and imaginary spin coherences, as follows from Eqs.~\ref{eq:spin-visible-polarization-components} and~\ref{eq:spin-regional-polarization}.

The two assignments place the $\Lambda$ in different current matrix elements. 
In assignment $23$, it is the daughter of the $\Lambda_b\to\Lambda$ transition. 
Assignment $13$ instead creates the $\Lambda$ through charged $\Lambda\bar p$ pair production and contains both vector-axial and scalar-pseudoscalar products. 
At the reference inputs, Table~\ref{table:spin-visible-regions} shows that the transition assignment produces an almost fully negative longitudinal polarization, whereas the charged-pair assignment has a larger fraction of positive-longitudinal-spin states and substantial transverse coherence.
The cut $m(p\bar p)<2.85\,\allowbreak \mathrm{GeV}$ strengthens the negative longitudinal polarization through two effects already visible before the assignments interfere. Assignment $23$, with $\mathcal P_L\simeq-1$, increases from about $25\%$ of the incoherent rate in full phase space to about $62\%$ after selection. At the same time, the longitudinal average of assignment $13$ changes from $-0.653$ to $-0.817$, and its transverse magnitude decreases from $0.350$ to $0.107$. 
The selected incoherent state therefore has $\mathcal P_L=-0.929$, compared with $-0.738$ in full phase space. 
The selection changes both the contribution mixture and its internal spin composition.
Coherence further shifts $\mathcal P_L$ to $-0.803$ in full phase space and $-0.954$ after selection. The mechanism follows from the signed numerator $\mathcal N_L=\mathcal B_\Lambda\mathcal P_L$ and the normalized-shift identity of Eq.~(\ref{eq:normalized-interference-shift}). 
Using the displayed values in Table~\ref{table:spin-visible-regions}, the longitudinal interference numerator is estimated as approximately $-1.86\times10^{-6}$ in the full region and $-1.15\times10^{-6}$ in the selected region, alongside positive rate interference of $1.90\times10^{-6}$ and $1.16\times10^{-6}$.
Thus constructive rate interference preferentially enhances the negative longitudinal spin projection. The coherent negative-spin fraction $(1-\mathcal P_L)/\allowbreak 2$ rises from about $90\%$ to $98\%$ under selection.

\begin{table}[htbp]
\centering\small
\caption{Reference visible-plane regional polarizations and particle branching fractions. Isolated rows retain one assignment, whereas the incoherent row sums their densities before normalization. Interference is a signed branching contribution. Dashes denote its undefined polarization.}
\label{table:spin-visible-regions}
\renewcommand{\arraystretch}{1.16}
\setlength{\tabcolsep}{5pt}
\begin{fitwidthtable}
\begin{tabular}{@{}llrrrr@{}}
\toprule
Region & Contribution & $10^6\mathcal B_\Lambda$ & $\mathcal P_L$ & $\mathcal P_T$ & $\mathcal P_N$ \tabularnewline
\midrule
Full & Assignment $13$ & $3.84$ & $-0.653$ & $-0.350$ & $0.00987$ \tabularnewline
Full & Assignment $23$ & $1.27$ & $-0.995$ & $0.0110$ & $0$ \tabularnewline
Full & Incoherent & $5.10$ & $-0.738$ & $-0.261$ & $0.00742$ \tabularnewline
Full & Interference & $1.90$ & --- & --- & --- \tabularnewline
Full & Coherent & $7.00$ & $-0.803$ & $-0.226$ & $0.00781$ \tabularnewline
Selected & Assignment $13$ & $0.703$ & $-0.817$ & $-0.107$ & $0.00511$ \tabularnewline
Selected & Assignment $23$ & $1.16$ & $-0.997$ & $0.0131$ & $0$ \tabularnewline
Selected & Incoherent & $1.86$ & $-0.929$ & $-0.0324$ & $0.00193$ \tabularnewline
Selected & Interference & $1.16$ & --- & --- & --- \tabularnewline
Selected & Coherent & $3.02$ & $-0.954$ & $-0.0458$ & $0.00369$ \tabularnewline
\bottomrule
\end{tabular}
\end{fitwidthtable}
\end{table}

\begin{figure}[tbp]
\centering
\includegraphics[width=\linewidth,height=0.5\textheight,keepaspectratio]{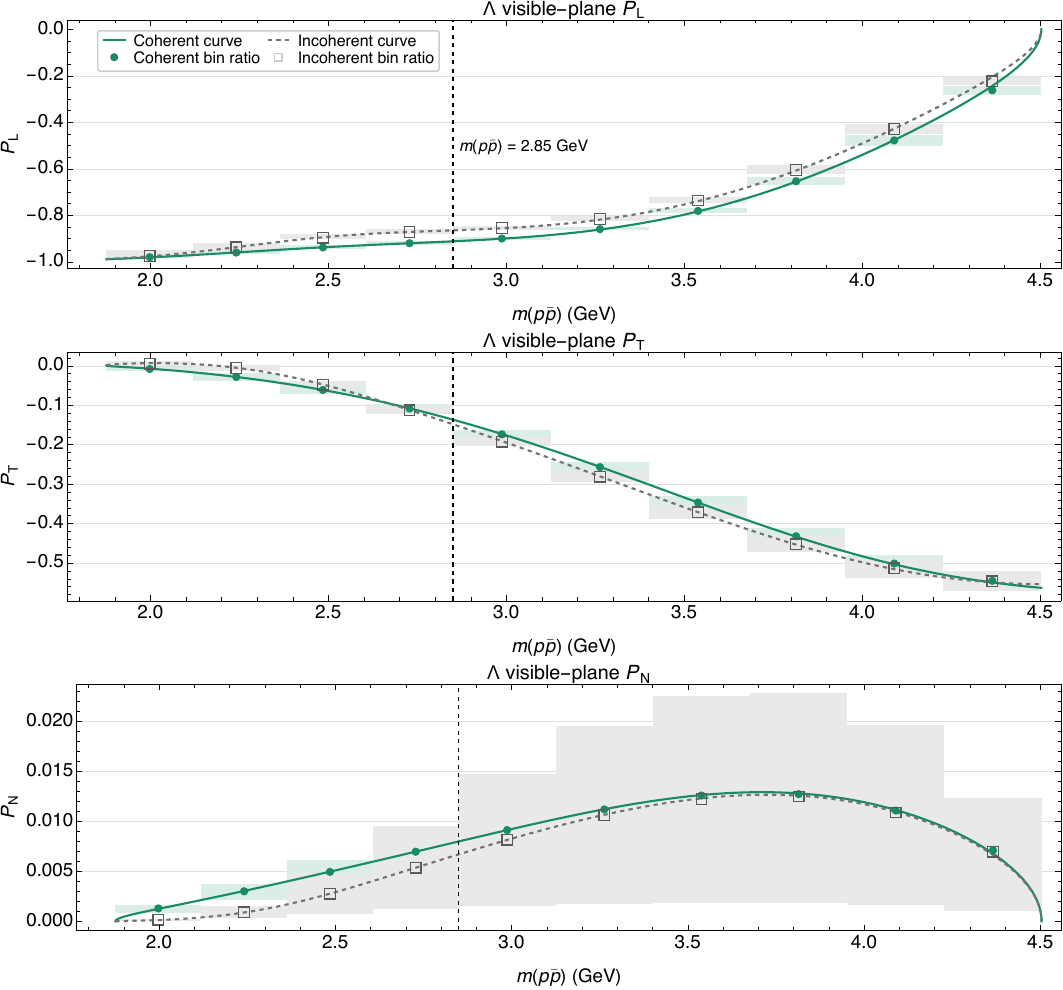}
\caption{ 
$\Lambda$ polarization in the visible-plane frame as a function
of $m(p\bar p)$ for coherent and incoherent contributions at the
reference inputs. 
Filled circles and open squares denote the coherent and incoherent rate-weighted bin ratios, respectively.
Rectangles span the mass bins and show nominal transition-only $16$th--$84$th percentile intervals.
The covariance prescription is specified in Section~\ref{sec:5-1}. 
The selection $m(p\bar p)<2.85\,\mathrm{GeV}$ is marked by a common bin boundary.
}
\label{fig:spin-visible-curves}
\end{figure}

\begin{figure}[tbp]
\centering
\includegraphics[width=\linewidth,height=0.5\textheight,keepaspectratio]{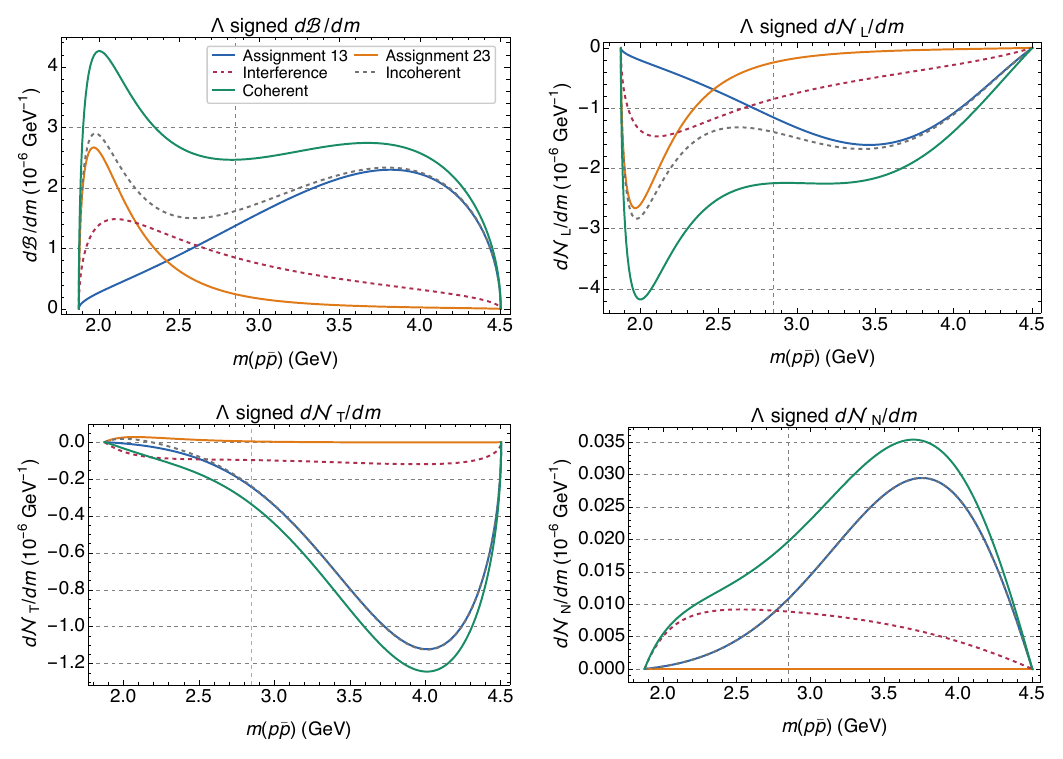}
\caption{
Assignment decomposition of the scalar branching-fraction and signed visible-spin-numerator spectra in $m(p\bar p)$ for $\Lambda_b^0\to\Lambda p\bar p$ at the reference inputs.
}
\label{fig:spin-signed-spectra}
\end{figure}

Fig.~\ref{fig:spin-visible-curves} and Fig.~\ref{fig:spin-signed-spectra} locate these changes in $m(p\bar p)$. The selected low-mass region carries a strongly negative longitudinal component, whereas the region above the cut has a different longitudinal and transverse composition.
The signed spin-numerator spectra show how each mass region contributes to the integrated spin numerator. Integrating that numerator and the rate over the selected region gives the corresponding regional polarization.
In the selected region, interference changes the normal polarization from $0.00193$ to $0.00369$. 
The latter is smaller than the full-region coherent value of $0.00781$. Absorptive or additional strong phases can generate this T-odd component, so nonzero normal polarization alone does not establish CP violation~\cite{GronauRosner2011}.
The selected longitudinal value is $\mathcal P_L=-0.9537 \pm0.0092_{\mathrm{stat},\Delta} \pm0.0049_{\mathrm{HO,partial},\Delta}$.
The uncertainties arise from the proton and $\Lambda$ transition-form-factor inputs, as described in Section~\ref{sec:5-1}.

\FloatBarrier
\subsection{CP phenomenology and experimental assessment}
\label{sec:5-6}

Direct rate CP violation requires interference of contributions with different weak and strong phases. 
The conjugate amplitude is constructed using the prescription of Section~4.6. 
The absorptive parts of $c_i^{\mathrm{eff}}$ supply strong phases even when the timelike form factors are real and the added relative phase is $\delta_\Lambda=0$.

\begin{figure}[tbp]
\centering
\includegraphics[width=\linewidth,height=0.5\textheight,keepaspectratio]{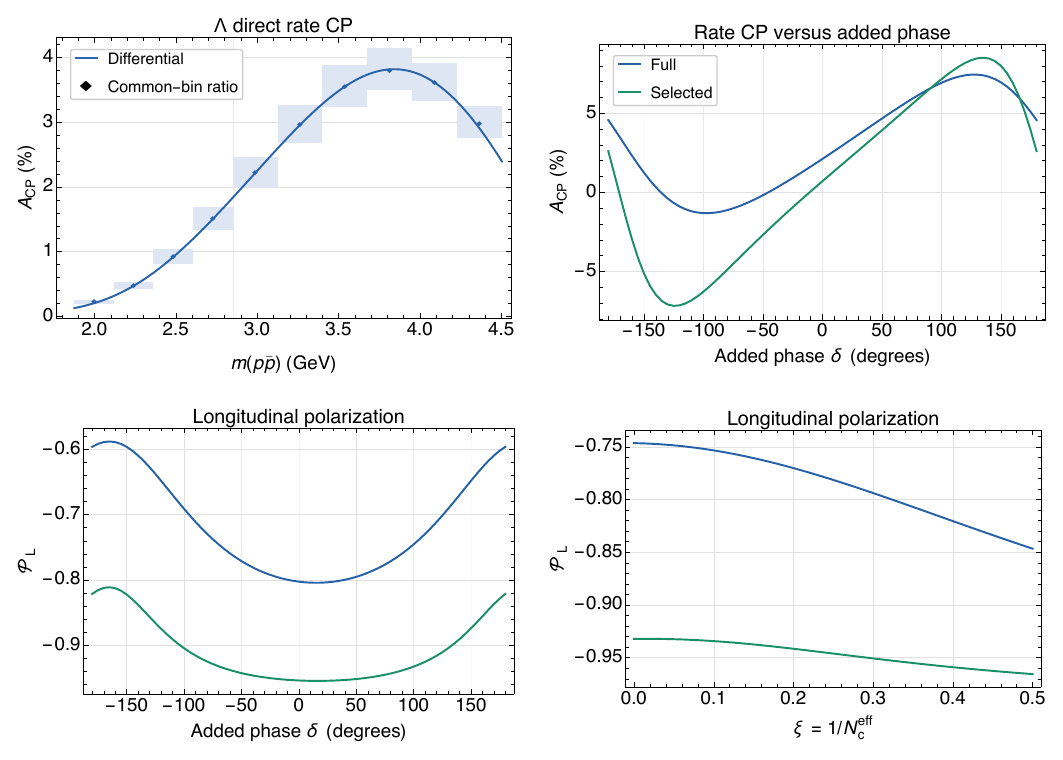}
\caption{
Rate CP asymmetries and longitudinal polarization in the $\Lambda$ channel.
The upper-left panel shows the differential and common-bin reference $A_{\mathrm{CP}}^\Lambda$ versus $m(p\bar p)$ at $N_c^{\mathrm{eff}}=3$ and $\delta_\Lambda=0$. 
Diamonds denote rate-weighted bin ratios. 
Rectangles span the mass bins and show nominal transition-only $16$th--$84$th percentile intervals.
The upper-right panel shows full-region and selected rate CP asymmetries versus the common CP-even phase at $N_c^{\mathrm{eff}}=3$. 
The lower panels show the full-region and selected longitudinal polarization $\mathcal P_L$ versus the added phase on the left and the inverse effective-color parameter at $\delta_\Lambda=0$ on the right.
Only the upper-left panel includes mass-bin intervals. 
The other panels show fixed-input phase or effective-color sensitivities.
}
\label{fig:spin-rate-cp-sensitivity}
\end{figure}

Tables~\ref{table:lambda-cp-results} and~\ref{table:spin-rate-cp-uncertainty} give the resulting direct asymmetry $A_{\mathrm{CP}}^\Lambda(\Omega)$ and CP average $\mathcal B_{\Lambda,\mathrm{av}}(\Omega)$. 
The reference full-region asymmetry is $+2.10\%$, with CP-averaged branching fraction $6.86\times10^{-6}$. 
After selection, these values become $+0.69\%$ and $3.00\times10^{-6}$.
The full-region asymmetry has the same sign as the earlier $+3.4\%$ central prediction in the same particle-minus-antiparticle convention~\cite{HsiaoGengRodrigues2019}. 
The different inputs and strong-phase prescriptions limit this comparison to the sign of the central values.
The change from $+2.10\%$ to $+0.69\%$ reflects the mass dependence of both the CP-odd branching difference $\mathcal B_\Lambda-\overline{\mathcal B}_\Lambda$ and its CP-even normalization $2\mathcal B_{\Lambda,\mathrm{av}}$.
At the reference inputs, the first low-mass bin has asymmetry $+0.21\%$, the highest selected bin has $+1.50\%$ and the complementary region above the cut has $+3.20\%$ (Fig.~\ref{fig:spin-rate-cp-sensitivity}). 
The full-region asymmetry is the average of the regional asymmetries below and above the cut, weighted by their CP-averaged partial branching fractions. 
The cut retains a CP-averaged signal fraction $0.4376\pm0.0373_{\mathrm{stat},\Delta}\pm0.0266_{\mathrm{HO,partial},\Delta}$ and removes the higher-asymmetry region, giving the selected value $+0.69\%$. 

\begin{table}[htbp]
\centering\small
\caption{Central $\Lambda$ particle, antiparticle and CP-averaged branching fractions $\mathcal B_\Lambda(\Omega),\allowbreak \overline{\mathcal B}_\Lambda(\Omega),\allowbreak \mathcal B_{\Lambda,\mathrm{av}}(\Omega)$, in units of $10^{-6}$, and direct CP asymmetry $A_{\mathrm{CP}}^\Lambda(\Omega)$, in percent. 
Both assignments and the specified four-quark operator content are included at $N_c^{\mathrm{eff}}=3$ with $\delta_\Lambda=0$.
The selected row imposes $m(p\bar p)<2.85\,\mathrm{GeV}$ on both conjugate modes. 
}
\label{table:lambda-cp-results}
\renewcommand{\arraystretch}{1.16}
\setlength{\tabcolsep}{5pt}
\begin{fitwidthtable}
\begin{tabular}{@{}lrrrr@{}}
\toprule
Region & $\mathcal B_\Lambda$ & $\overline{\mathcal B}_\Lambda$ & $\mathcal B_{\Lambda,\mathrm{av}}$ & $A_{\mathrm{CP}}^\Lambda\,\allowbreak [\%]$ \tabularnewline
\midrule
Full & $7.00$ & $6.71$ & $6.86$ & $2.10$ \tabularnewline
Selected & $3.02$ & $2.98$ & $3.00$ & $0.693$ \tabularnewline
\bottomrule
\end{tabular}
\end{fitwidthtable}
\end{table}

\begin{table}[htbp]
\centering\small
\caption{Correlated partial joint-transition errors for the retained CP average and rate asymmetry. Entries give the central value, statistical delta error and partial higher-order estimate. Asymmetry errors are percentage points. Both charges share the varied fit parameters.}
\label{table:spin-rate-cp-uncertainty}
\renewcommand{\arraystretch}{1.16}
\setlength{\tabcolsep}{5pt}
\begin{fitwidthtable}
\begin{tabular}{@{}lrr@{}}
\toprule
Region & \shortstack[l]{$10^6\mathcal B_{\Lambda,\mathrm{av}}$: central\\$\pm\sigma_{\mathrm{stat},\Delta}\pm\sigma_{\mathrm{HO,partial},\Delta}$} & \shortstack[l]{$A_{\mathrm{CP}}^\Lambda\,\allowbreak [\%]$: central\\$\pm\sigma_{\mathrm{stat},\Delta}\pm\sigma_{\mathrm{HO,partial},\Delta}$} \tabularnewline
\midrule
Full & $6.86\pm0.85\pm2.22$ & $2.10\pm0.23\pm0.19$ \tabularnewline
Selected & $3.00\pm0.33\pm0.73$ & $0.69\pm0.10\pm0.09$ \tabularnewline
\bottomrule
\end{tabular}
\end{fitwidthtable}
\end{table}

The dependence of the predicted asymmetry on the effective-color parameter and the additional relative phase is examined through one-dimensional scans with fixed hadronic inputs. 
At $\delta_\Lambda=0$, the choices $N_c^{\mathrm{eff}}=2,\allowbreak 3,\allowbreak 4,\allowbreak 5,\allowbreak \infty$ give full-region asymmetries $1.13\%,\allowbreak 2.10\%,\allowbreak 2.91\%,\allowbreak 3.50\%,\allowbreak 6.32\%$ and selected asymmetries $0.160\%,\allowbreak 0.693\%,\allowbreak 1.62\%,\allowbreak 2.49\%,\allowbreak 7.69\%$. 
At $N_c^{\mathrm{eff}}=3$, the $73$-point phase scan over $\delta_\Lambda\in[-\pi,\allowbreak \pi]$ gives sampled ranges $[-1.32,\allowbreak +7.44]\%$ and $[-7.19,\allowbreak +8.50]\%$, respectively. 
Fig.~\ref{fig:spin-rate-cp-sensitivity} compares the mass-resolved rate asymmetry with the phase and color responses of the longitudinal polarization. 
These ranges show the fixed-input response to each varied parameter.

The selected CP-averaged rate can now be compared with experiment, provided the signal and normalization selections are matched. 
LHCb measures $R_{\mathrm{cut}}=(5.1\pm1.3_{\mathrm{stat}}\pm0.3_{\mathrm{syst}})\times10^{-2}$ with $m(p\bar p)<2.85\,\mathrm{GeV}$ for the signal and $m(K^+K^-)<2.85\,\mathrm{GeV}$ for the normalization mode $\Lambda_b^0\to\Lambda K^+K^-$~\cite{LHCbLambdaPPbar2026}. 
The available absolute normalization, $\mathcal B_{KK}^{\mathrm{full}}=(15.9\pm1.2_{\mathrm{stat}}\pm1.2_{\mathrm{syst}}\pm2.0_{\mathrm{norm}})\times10^{-6}$, has no additional kaon-pair mass cut~\cite{LHCbLambdaHH2016}. 
With compatible charge and charm definitions, the normalization fraction $f_{KK}$ below the cut relates the measured ratio to the selected signal branching fraction:
\begin{equation}
\label{eq:experimental-cut-conversion}
\begin{aligned}
f_{KK}
&=\frac{\mathcal B_{KK}^{\mathrm{cut}}}
{\mathcal B_{KK}^{\mathrm{full}}},\qquad 0\leq f_{KK}\leq1,\\
\mathcal B_{p\bar p}^{\mathrm{cut}}
&=R_{\mathrm{cut}}\,f_{KK}\,\mathcal B_{KK}^{\mathrm{full}}
\leq R_{\mathrm{cut}}\,\mathcal B_{KK}^{\mathrm{full}}.
\end{aligned}
\end{equation}
At the experimental central inputs, the available full-region normalization gives an upper envelope of $0.811\times10^{-6}$ for the selected signal branching fraction. 
The reference selected CP average of $3.00\times10^{-6}$ lies above this envelope. 
It would require $f_{KK}\simeq3.70$ under the fixed central inputs, outside the physical interval. 
This comparison identifies a difference at the reference point and does not by itself establish a statistical inconsistency with experiment.

\begin{figure}[tbp]
\centering
\includegraphics[width=\linewidth,height=0.5\textheight,keepaspectratio]{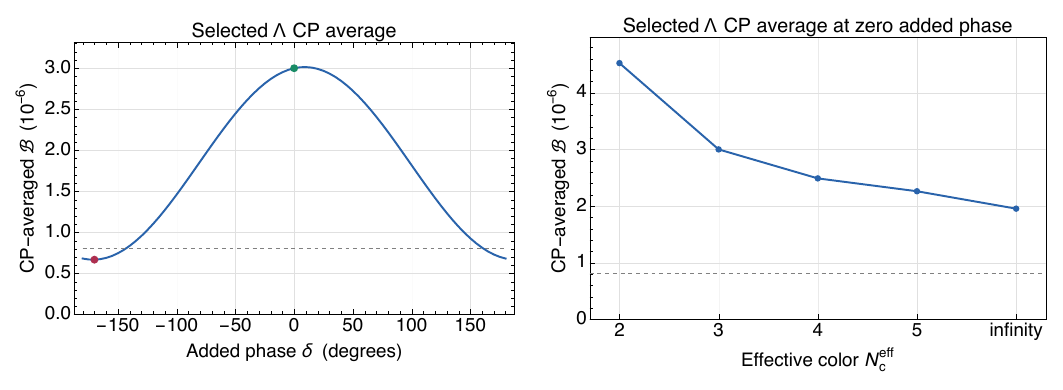}
\caption{Selected $\Lambda$ CP-averaged branching fraction $10^6\mathcal B_{\Lambda,\mathrm{av}}(\Omega)$ with $m(p\bar p)<2.85\,\allowbreak \mathrm{GeV}$ on both charges.
Left: added-phase scan at $N_c^{\mathrm{eff}}=3$, with the reference and lowest sampled point marked. 
Right: $N_c^{\mathrm{eff}}=2,\allowbreak 3,\allowbreak 4,\allowbreak 5,\allowbreak \infty$ at $\delta_\Lambda=0$. 
The same CP-even phase is used for both conjugate amplitudes. 
The dashed line marks $R_{\mathrm{cut}}\mathcal B_{KK}^{\mathrm{full}}=0.811\times10^{-6}$ at fixed experimental central inputs, the envelope obtained from $0\leq f_{KK}\leq1$.}
\label{fig:cp-average-sensitivity}
\end{figure}

However, Fig.~\ref{fig:cp-average-sensitivity} shows how this comparison depends on the relative current phase. At zero added phase, all five effective-color choices remain above the envelope. At $N_c^{\mathrm{eff}}=3$, the sampled selected CP average spans $[0.669,\allowbreak 3.014]\times10^{-6}$.
The sampled point near $\delta_\Lambda=-170^\circ$ gives $0.669\times10^{-6}$, corresponding to the required fraction $f_{KK}^{\mathrm{required}}\simeq0.825$. 
This point satisfies a necessary central-value condition but does not by itself establish agreement with experiment. It is a different parameter point from the zero-phase benchmark used for the polarization values in Section~5.5.

\FloatBarrier
\section{Conclusions}
\label{sec:6}
An angular and spin analysis of the charmless purely baryonic decays $\Lambda_b^0\to n p\bar p$ and $\Lambda_b^0\to\Lambda p\bar p$ has been carried out within a phenomenological generalized-factorization framework. Two current contributions to each final state are combined coherently. The hadronic inputs consist of lattice-QCD transition form factors and phenomenological timelike baryon-pair form factors. A common external-spin representation is used to calculate pair-frame angular distributions, hemisphere asymmetries, and Legendre moments in both channels, together with the $\Lambda$ polarization for an unpolarized parent.

In the neutron channel, the hemisphere asymmetry at $N_c^{\mathrm{eff}}=3$ and zero additional relative phase is $4.46\times10^{-4}$ in the $n\bar p$ rest frame, compared with $0.856$ in the $p\bar p$ rest frame. Invariant-mass binning shows that the small first value results from cancellation between negative lower-mass and positive higher-mass contributions to the forward-minus-backward rate. Thus, a small integrated asymmetry need not imply a nearly symmetric distribution throughout phase space. Appreciable regional asymmetries can be hidden by the integration. The difference between direct hemisphere integration and the first-moment estimate also illustrates the role of higher odd angular moments.

In the $\Lambda$ channel, the selection $m(p\bar p)<2.85\,\mathrm{GeV}$ preferentially removes the backward part of the crossed $\Lambda\bar p$ angular domain. At the same reference inputs, the rate-weighted longitudinal polarization along the $\Lambda$ flight direction changes from $-0.803$ to $-0.954$. This change reflects both the altered relative weights of the current contributions and their individual spin compositions. 
Constructive interference further strengthens the preference for the negative longitudinal spin orientation.
The result shows that an invariant-mass selection can change the observed spin composition as well as the accepted rate, so the polarization must be interpreted together with the selected phase-space region.

With the same invariant-mass selection imposed on the decay and its CP-conjugate mode, the predicted direct CP asymmetry decreases from $+2.10\%$ to $+0.69\%$, because the selection removes a region with a larger rate asymmetry. The comparison with experiment also depends on the normalization-channel branching fraction in the corresponding selected phase-space region. Using the available full-region normalization and the experimental central inputs gives an upper envelope of $0.811\times10^{-6}$ for the selected signal branching fraction. The reference CP-averaged prediction, $3.00\times10^{-6}$, lies above this envelope, whereas some sampled relative phases satisfy the corresponding necessary central-value condition. 
These angular, spin, and phase dependences are obtained within the adopted generalized-factorization framework and should be interpreted in that context. In particular, nonfactorizable strong-interaction effects are not fully accounted for. The results nevertheless provide a basis for further theoretical studies and for future measurements of angular and polarization observables in these decays.

\section*{Acknowledgments} 
This work is supported by 
the National Natural Science Foundation of China (NSFC) under Grant No. 12547104 and the China Postdoctoral Science Foundation under Grant No. 2026M793678.

\clearpage

\appendix

\FloatBarrier

\section{Operator conventions and input specification}

This appendix collects the operator conventions and hadronic inputs used in Section~2. It first specifies the four-quark operator basis and the current-density relations used for the charged-pair diagnostic. It then gives the conversion between lattice helicity and covariant transition form factors, the timelike flavor and weak coefficients, and the lattice-fit parameterization.

\subsection{Operator basis and current-density relations}

The chiral notation is $(\bar a b)_{V\pm A}=\bar a\gamma^\mu(1\pm\gamma_5)b$, with Lorentz indices contracted between the two bilinears. In the color convention used to define the effective coefficients $a_i$ in Section~2.1, the four-quark operators are~\cite{GengLiuTsai2021}
$$
\begin{aligned}
O_1^{r,q}
&=
(\bar q_i r_i)_{V-A}
(\bar r_j b_j)_{V-A},
&
O_2^{r,q}
&=
(\bar q_i r_j)_{V-A}
(\bar r_j b_i)_{V-A},\\
O_{3,5}^{(q)}
&=
(\bar q_i b_i)_{V-A}
\sum_f(\bar f_j f_j)_{V\mp A},
&
O_{4,6}^{(q)}
&=
(\bar q_i b_j)_{V-A}
\sum_f(\bar f_j f_i)_{V\mp A},\\
O_{7,9}^{(q)}
&=
\frac{3}{2}
(\bar q_i b_i)_{V-A}
\sum_f e_f(\bar f_j f_j)_{V\pm A},
&
O_{8,10}^{(q)}
&=
\frac{3}{2}
(\bar q_i b_j)_{V-A}
\sum_f e_f(\bar f_j f_i)_{V\pm A}.
\end{aligned}
$$
The upper sign refers to the first index in each pair of subscripts. The flavor sum in the Hamiltonian runs over active quarks, while the explicit charmless pair currents considered in the amplitudes contain $u$, $d$, and $s$ quarks. The phenomenological combination of the two current products is specified in Section~2.2.
The left-right penguin operators give rise to scalar-pseudoscalar products through the four-dimensional Fierz identity,
$$
\left(\bar\psi_1\gamma^\mu(1-\gamma_5)\psi_2\right)
\left(\bar\psi_3\gamma_\mu(1+\gamma_5)\psi_4\right)
=
-2
\left(\bar\psi_1(1+\gamma_5)\psi_4\right)
\left(\bar\psi_3(1-\gamma_5)\psi_2\right).
$$
The minus sign combines with the sign of the penguin terms in the Hamiltonian to give the scalar coefficients $\kappa_q^{SP}$ and $\eta_{f,q}$ in Eqs.~\eqref{eq:unified-charged-amplitude} and \eqref{eq:unified-neutral-flavor-map}. 
The chiral densities use the matrix-element definitions of Section~2.3 without an additional Fierz factor absorbed into them. 
The four-dimensional reduction is used with the adopted effective-coefficient prescription. 
Higher-order operator matching, including evanescent contributions, is not recalculated here~\cite{AebischerBurasKumar2023}.

The reference calculation uses the fitted timelike scalar and pseudoscalar pair-density functions described in Section~\ref{sec:2-3}. To examine the dependence on this input choice, a separate diagnostic reconstructs the charged-pair densities from nonsinglet current-density relations within the adopted current basis. The reconstructed functions are
$$
\begin{aligned}
F_{S,\mathrm W}^{qu}(t)
&=
\frac{m_{B_q}-m_p}{m_q-m_u}F_1^{qu}(t),\\
G_{P,\mathrm W}^{qu}(t)
&=
\frac{
(m_{\mathbf{B}_q}+m_p)g_A^{qu}(t)
+t\,h_A^{qu}(t)/(m_{\mathbf{B}_q}+m_p)
}
{m_q+m_u}.
\end{aligned}
$$
For the present current basis and inputs, the fitted pair densities do not jointly satisfy these relations. 

The pair-current parameterization in Eq.~\eqref{eq:unified-pair-currents} omits an allowed induced-scalar vector covariant for an unequal-mass flavor-changing pair. This covariant would contribute to the vector divergence and the scalar relation. The parameterization also omits the induced-tensor axial covariant, whose divergence vanishes. These omissions are assumptions of the adopted parameterization rather than consequences of the Ward identities.

In the neutral sector, vector-current conservation leaves the scalar density undetermined. The singlet component of the flavor-diagonal axial currents introduces a gluonic anomaly~\cite{Larin1993}, whose matrix element is also required for pseudoscalar reconstruction. The charged-pair diagnostic therefore does not determine a unique complete replacement of the reference pair inputs.

\subsection{Helicity-to-covariant conversion and endpoints} 

The lattice helicity form factors $f_+$, $f_\perp$, $f_0$, $g_+$, $g_\perp$, and $g_0$ are converted to the covariant transition form factors of Section~2.3 using $s_\pm=(M\pm m_N)^2-t$. In the positive-tensor convention, the conversion gives~\cite{DetmoldLehnerMeinel2015,DetmoldMeinel2016}
\begin{equation}
\label{eq:unified-lattice-conversion}
\begin{aligned}
f_1&=\frac{(M+m_N)^2f_+-tf_\perp}{s_+},
&
g_1&=\frac{(M-m_N)^2g_+-tg_\perp}{s_-},\\
f_2&=\frac{M(M+m_N)(f_+-f_\perp)}{s_+},
&
g_2&=\frac{M(M-m_N)(g_\perp-g_+)}{s_-},\\
f_3&=\frac{M(M-m_N)(f_0-f_1)}{t},
&
g_3&=\frac{M(M+m_N)(g_1-g_0)}{t}.
\end{aligned}
\end{equation}
The current divergences give $f_S^N=(M-m_N)f_0/(m_b-m_{q_N})$ and $g_P^N=(M+m_N)g_0/(m_b+m_{q_N})$. 
The transition density form factors are obtained from the converted helicity functions through these relations.

At zero recoil, the shared axial intercept enforces $g_\perp=g_+$~\cite{HillerZwicky2021}. The corresponding divided difference is evaluated analytically to avoid a numerical $0/0$. At zero momentum transfer, the official fits impose soft priors rather than every exact equality at $t=0$. The conversion is used only in the physical pair region $t>0$, so the factors of $1/t$ do not require evaluation at zero momentum transfer. The conformal-map thresholds and expansion points are specified in Section~A.4.

For the neutron, the proton helicity functions are first converted to covariant form factors. They are then transferred using $f_i^n=f_i^p$ and $g_i^n=g_i^p$, with a fixed positive relative isospin phase. The functions are evaluated at the current invariant appropriate to each assignment. The $\Lambda$ transition uses its separately released fit.


\subsection{Timelike flavor and weak coefficients}

With the current definitions of Section~2.3, the charged $\Lambda$ coefficients follow the signs in Eqs. (10) and (11) of Ref.~\cite{HuangHsiaoWangSun2022}. They are
$$
\begin{aligned}
C_{F_1}^{su}
&=
\sqrt{\frac32}C_\parallel,
&
C_{g_A}^{su}
&=
\sqrt{\frac32}
\left(C_\parallel+\delta C_\parallel\right),\\
C_{F_S}^{su}
&=
-\sqrt{\frac32}\bar C_\parallel,
&
C_{G_P}^{su}
&=
-\sqrt{\frac32}
\left(\bar C_\parallel+\delta\bar C_\parallel\right),\\
C_{h_A}^{su}
&=
-\frac{C_D+3C_F}{\sqrt6}.
\end{aligned}
$$
The charged flavor prefactors printed in Refs.~\cite{HsiaoGeng2015,HuangHsiaoWangSun2022} differ, but this difference alone does not establish opposite physical vector or axial signs. 
Their fitted constants also differ, and both central prescriptions give positive charged coefficients. The induced-axial functions use different powers and dimensions, so they cannot be reconciled by a common external-state phase. 
The neutral prescription $C_{h_A}^{dd}=C_F-C_D$ follows the strange-current omission stated in Section~2.3. The value $-(C_D+C_F)$ printed in Ref.~\cite{Hsiao2023} remains a separate sensitivity choice. The supporting numerical input records contain the detailed source comparison and the derived vector, axial, and density coefficients.

The pair constants used in these coefficients are taken from the global fit of Ref.~\cite{HuangHsiaoWangSun2022}. 
The neutral induced-axial coefficients follow the prescription specified in Section~\ref{sec:5-1}. 
The CKM benchmark and quark masses follow Ref.~\cite{GengLiuTsai2021}.
The neutron effective coefficients at $\mu=2.5\,\mathrm{GeV}$ use the NDR and electromagnetic conventions stated in Section~\ref{sec:5-1}~\cite{GengLiuTsai2021}. 
They are
\begin{equation}
\label{eq:central-effective-coefficients}
\begin{aligned}
(c_1^{\mathrm{eff}},c_2^{\mathrm{eff}})
&=(1.168,-0.365),\\
(c_3^{\mathrm{eff}},c_4^{\mathrm{eff}})
&=(0.0238+0.0014i,-0.0497-0.0042i),\\
(c_5^{\mathrm{eff}},c_6^{\mathrm{eff}})
&=(0.0145+0.0014i,-0.0633-0.0042i),\\
(c_7^{\mathrm{eff}},c_8^{\mathrm{eff}})
&=(-0.00010-0.00010i,0.00050),\\
(c_9^{\mathrm{eff}},c_{10}^{\mathrm{eff}})
&=(-0.01120-0.00013i,0.00200).
\end{aligned}
\end{equation}
The current-current coefficients $c_1^{\mathrm{eff}}$ and $c_2^{\mathrm{eff}}$ also apply to the $\Lambda$ channel. 
Its remaining coefficients are given in Table~\ref{table:lambda-effective-coefficients}. 
The antiparticle column is used together with conjugated CKM products.

\begin{table}[htbp]
\centering\small
\caption{
Effective coefficients for the $\Lambda$ channel at $\mu=2.5\,\mathrm{GeV}$~\cite{GengLiuTsai2021}. 
The entries are $10^4c_i^{\mathrm{eff}}$. 
The antiparticle column is used together with conjugated CKM products.
}
\label{table:lambda-effective-coefficients}
\renewcommand{\arraystretch}{1.16}
\setlength{\tabcolsep}{5pt}
\begin{fitwidthtable}
\begin{tabular}{@{}lrr@{}}
\toprule
Index $i$ & $b\to s$ & $\bar b\to\bar s$ \tabularnewline
\midrule
$3$ & $243+31i$ & $241+32i$ \tabularnewline
$4$ & $-512-94i$ & $-506-97i$ \tabularnewline
$5$ & $150+31i$ & $148+32i$ \tabularnewline
$6$ & $-649-94i$ & $-643-97i$ \tabularnewline
$7$ & $-1.1-2.2i$ & $-1.1-1.3i$ \tabularnewline
$8$ & $5$ & $5$ \tabularnewline
$9$ & $-112-2.2i$ & $-112-2.2i$ \tabularnewline
$10$ & $20$ & $20$ \tabularnewline
\bottomrule
\end{tabular}
\end{fitwidthtable}
\end{table}

\subsection{Released lattice fits}

The lattice helicity form factors use the parameterization~\cite{DetmoldLehnerMeinel2015,DetmoldMeinel2016}
$$
F(t)
=
\frac{a_0+a_1z(t)+a_2z(t)^2}
{1-t/m_{\mathrm{pole}}^2},
\qquad
z(t)
=
\frac{
\sqrt{t_+-t}-\sqrt{t_+-t_0}
}{
\sqrt{t_+-t}+\sqrt{t_+-t_0}
}.
$$
For the proton transition, the threshold and expansion point are $t_+=(5.279+0.1348)^2\,\mathrm{GeV}^2$ and $t_0=(M-m_p)^2$. For the $\Lambda$ transition, they are $t_+=(5.279+0.494)^2\,\mathrm{GeV}^2$ and $t_0=(M-m_\Lambda)^2$.
The external baryon masses are given in Eq.~\eqref{eq:central-input-values}. 
The published coefficients, covariance matrices, pole masses and branch thresholds are used without refitting. 
For the neutron, the converted proton covariant form factors are evaluated at the invariant mass of each current assignment.

The nominal fit has $a_2=0$, while the higher-order fit includes the quadratic term. 
Both sources impose the shared axial intercept exactly, allowing the finite zero-recoil divided difference to be evaluated analytically. 
The conversion follows the endpoint prescription in Section~A.2. Rounded nominal tables are retained in the supporting records for identification. 
The numerical inputs are specified by the unchanged released coefficients and covariance in the source package.


\section{Tensor representation and kernel construction}

This appendix makes the bilinear construction of Section~3 explicit. The current-product columns and coefficient signs specify the matrix representation, and connected traces provide independent checks of its interference terms. The indexed retained-spin contraction then fixes the parent-density transpose and normalization. 

\FloatBarrier
\subsection{Explicit current-product columns}

Factoring form factors and weak coefficients out of the current-product matrix $X$ leaves kinematic Dirac structures. For each assignment's current momentum $q$, the six transition matrices are
\begin{equation}
\label{eq:unified-transition-tensors}
\begin{aligned}
A_1^\mu&=\gamma^\mu,&
A_2^\mu&=\frac{i\sigma^{\mu\nu}q_\nu}{M},&
A_3^\mu&=\frac{q^\mu}{M}\mathbf1_4,\\
A_4^\mu&=\gamma^\mu\gamma_5,&
A_5^\mu&=\frac{i\sigma^{\mu\nu}q_\nu}{M}\gamma_5,&
A_6^\mu&=\frac{q^\mu}{M}\gamma_5.
\end{aligned}
\end{equation}
The pair matrices are $B_j^\mu=(\gamma^\mu,\allowbreak i\sigma^{\mu\nu}q_\nu/\allowbreak \mu_a,\allowbreak \gamma^\mu\gamma_5,\allowbreak q^\mu\gamma_5/\allowbreak \mu_a)$, with $\mu_{13}=m_Y+m_p$, $\mu_{23}=2m_p$, and density matrices $C_S=\mathbf1_4$, $C_P=\gamma_5$. With $u_0=u_{\Lambda_b}$, the columns are
\begin{equation}
\label{eq:unified-tensor-columns}
\begin{aligned}
X^{VA}_{h,13;ij}
&=(\bar u_2A_i^\mu u_0)(\bar u_1B_{j,\mu}v_3),\quad
X^{VA}_{h,23;ij}
=(\bar u_1A_i^\mu u_0)(\bar u_2B_{j,\mu}v_3),\\
X^{SP}_{h,13;rs}
&=(\bar u_2C_ru_0)(\bar u_1C_sv_3),\quad
X^{SP}_{h,23;rs}
=(\bar u_1C_ru_0)(\bar u_2C_sv_3).
\end{aligned}
\end{equation}
Each assignment uses its own momentum. Vector-axial columns run first over the transition index and then the pair index, while density columns follow $(SS,\allowbreak SP,\allowbreak PS,\allowbreak PP)$. 
The full order is $(13\,\allowbreak VA[24],\allowbreak 13\,\allowbreak SP[4],\allowbreak 23\,\allowbreak VA[24],\allowbreak 23\,\allowbreak SP[4])$. 
The sixteen external-spin configurations give a $16\times56$ matrix $X$ of redundant current-product coordinates, rather than 56 independent physical helicity amplitudes. 
The pointwise Gram rank is at most sixteen. 
However, the integrated response $R[w]$ need not share this bound because null directions can vary across phase space.

The signed transition, charged-pair and density coefficient vectors are
\begin{equation}
\label{eq:scalar-coefficient-order}
\begin{aligned}
&\boldsymbol v_{\mathrm{tr}}=(f_1,f_2,f_3,-g_1,-g_2,-g_3),\quad
\boldsymbol v_{\mathrm{pair}}=(F_1,F_2,-g_A,-h_A),\\
&\boldsymbol v_{SP}=(f_SF_S,f_SG_P,-g_PF_S,-g_PG_P).
\end{aligned}
\end{equation}
The product $\boldsymbol v_{\mathrm{tr}}\otimes\boldsymbol v_{\mathrm{pair}}$ multiplies the charged vector-axial columns, and $\boldsymbol v_{SP}=(f_S,\allowbreak -g_P)\otimes(F_S,\allowbreak G_P)$ multiplies the density columns. 
Neutral vector and axial flavor coefficients $\beta_{f,q}^{+},\allowbreak \beta_{f,q}^{-}$ act separately on their pair subblocks. 
The signs preserve the axial subtraction and the pseudoscalar convention without an extra $i$.

The adopted neutral up- and down-vector inputs have the same proportionality in their Dirac components $F_1^{uu},\allowbreak F_1^{dd}$ and Pauli components $F_2^{uu},\allowbreak F_2^{dd}$. 
The independent weak-current combinations consequently decrease from seven to six for the neutron and, with the OZI approximation, from six to five for the $\Lambda$. 
Neutral density columns remain in $X$, with vanishing $\Lambda$ coefficients. 
These input-dependent reductions leave the general operator basis intact.

\FloatBarrier
\subsection{Connected traces and algebraic checks}

For relativistically normalized spinors, $\sum_su(p,\allowbreak s)\bar u(p,\allowbreak s)=\not p+m$ and $\sum_sv(p,\allowbreak s)\bar v(p,\allowbreak s)=\not p-m$. Write an assignment-$13$ column as $\sum_\ell w_\ell(\bar u_2U_{r\ell}u_0)(\bar u_1V_{r\ell}v_3)$, and its assignment-$23$ counterpart with weights $z_m$ and matrices $U_{sm},\allowbreak V_{sm}$. Set $S_0=\not P+M$, $S_1=\not p_1+m_Y$, $S_2=\not p_2+m_p$, $S_3=\not p_3-m_p$ and $\overline U=\gamma^0U^\dagger\gamma^0$. The connected trace gives
\begin{equation}
\label{eq:unified-cross-trace}
K_{13;r,23;s}
=\frac12\sum_{\ell,m}w_\ell^*z_m
\operatorname{Tr}\!\left[
S_0\overline U_{r\ell}S_2V_{sm}
S_3\overline V_{r\ell}S_1U_{sm}
\right].
\end{equation}
Lorentz contractions supply metric weights, whereas scalar products have unit weight. Diagonal Gram blocks reduce to products of transition and pair traces, while the connected trace retains the exchanged spinor lines, including vector-scalar cross terms. 

Opposite total-parity structures have zero unpolarized overlap because the momentum pseudoscalar $\epsilon_{\mu\nu\rho\sigma}P^\mu p_1^\nu p_2^\rho p_3^\sigma$ vanishes for $P=p_1+p_2+p_3$. Polarization introduces additional vectors and requires its own construction. The mass dimensions of $X$, $K$ and $H$ are two, four and minus two, respectively, giving a dimensionless invariant amplitude.

Direct current contraction, explicit spin sums and connected traces check tensor and pseudoscalar signs, antiparticle completeness and interference. A common unitary spin-basis change preserves $X^\dagger X$. A separately specified helicity chain also requires its rotations and phases. These algebraic and numerical checks assess consistency within the adopted amplitude.

\FloatBarrier
\subsection{Retained-spin contraction and parent-density transpose}

For the external-spin order $h=(s_b,\allowbreak s_\Lambda,\allowbreak \eta)$ and amplitude $\mathcal M_{s_b s_\Lambda\eta}=\sum_I X_{s_b s_\Lambda\eta,I}H_I$, the kernel associated with the physical Pauli response $W_\mu=\operatorname{Tr}(\rho_\Lambda\Sigma_\mu^\Lambda)$ is
\begin{equation}
\label{eq:spin-indexed-kernel}
[K_\mu[\rho_b]]_{IJ}
=\sum_{s_b,s'_b,s_\Lambda,s'_\Lambda,\eta}
X^*_{s'_b s'_\Lambda\eta,I}
(\rho_b)_{s_b s'_b}
(\Sigma_\mu^\Lambda)_{s'_\Lambda s_\Lambda}
X_{s_b s_\Lambda\eta,J}.
\end{equation}
The left row of $X$ carries primed indices and therefore requires $(\rho_b^T)_{s'_b s_b}=(\rho_b)_{s_b s'_b}$. The daughter operator already has the order $(\Sigma_\mu^\Lambda)_{s'_\Lambda s_\Lambda}$. This yields Eq.~(\ref{eq:spin-inserted-kernels}). The parent transpose follows from amplitude-vector ordering and leaves the physical parent preparation unchanged.

With daughter rows and parent columns, the amplitude matrix $\mathsf A_\eta$ gives
\begin{equation}
\label{eq:spin-matrix-density-check}
[\mathsf A_\eta\rho_b\mathsf A_\eta^\dagger]_{s_\Lambda s'_\Lambda}
=\sum_{s_b,s'_b}
\mathcal M_{s_b s_\Lambda\eta}
(\rho_b)_{s_b s'_b}
\mathcal M^*_{s'_b s'_\Lambda\eta}.
\end{equation}
Summing over $\eta$ gives $\rho_\Lambda$. The amplitude-matrix and flattened-kernel contractions can be compared for arbitrary physical $\rho_b$, including complex parent coherence. By contrast, unpolarized or real diagonal tests cannot detect a missing transpose.

For $\rho_b=\mathbf1/\allowbreak 2$ and $\Sigma_0^\Lambda=\mathbf1$, the scalar kernel is $K_0[\rho_b]=K=X^\dagger X/\allowbreak 2$, with exactly one parent average. A physical $\rho_b^T$ remains Hermitian and positive semidefinite, ensuring scalar-kernel positivity. The eigenvalues $\pm1$ of $\Sigma_i^\Lambda$ give Hermitian signed kernels and $|W_i|\leq W_0$. Oriented integration and axis transport require checks beyond these finite-dimensional identities.

\FloatBarrier

\clearpage
\begingroup
\small
\bibliographystyle{unsrt}
\bibliography{references}
\endgroup
\end{document}